\documentclass[twocolumn,footinbib,floatfix,noeprint,longbibliography, prb, superscriptaddress]{revtex4-2}

\usepackage{graphicx}
\usepackage{amsmath,amsfonts,amssymb}
\usepackage{mathtools}
\usepackage[colorlinks=true, citecolor=blue, linkcolor=red, urlcolor=blue]{hyperref}
\usepackage[all]{hypcap}
\usepackage{hhline}

\graphicspath{{figures/}}

\usepackage{bm}
\usepackage{commath}    
\usepackage{braket}

\newcommand{\bk}{{\bf k}}
\newcommand{\bK}{{\bf K}}
\newcommand{\bq}{{\bf q}}

\newcommand{\bA}{{\bf A}}

\newcommand{\bp}{{\bf p}}
\newcommand{\br}{{\bf r}}

\begin{document}

\title{Topological Density Correlations in a Fermi Gas}

\author{Pok Man Tam}
\affiliation{Department of Physics and Astronomy, University of Pennsylvania, Philadelphia PA 19104}
\affiliation{Princeton Center for Theoretical Science, Princeton University, Princeton, NJ 08544}

\author{Charles L. Kane}
\affiliation{Department of Physics and Astronomy, University of Pennsylvania, Philadelphia PA 19104}

\date{\today}

\begin{abstract}
A Fermi gas of non-interacting electrons, or ultra-cold fermionic atoms, has a quantum ground state defined by a region of occupancy in momentum space known as the Fermi sea. The Euler characteristic $\chi_F$ of the Fermi sea serves to topologically classify these gapless fermionic states. The topology of a $D$ dimensional Fermi sea is physically encoded in the $D+1$ point equal time density correlation function.  In this work, we first present a simple proof of this fact by 
showing that the evaluation of the correlation function can be formulated in terms of a triangulation of the Fermi sea with a collection of points, links and triangles and their higher dimensional analogs. We then make use of the topological $D+1$ point density correlation to reveal universal structures of the more general $M$ point density correlation functions in a $D$ dimensional Fermi gas. Two experimental methods are proposed for observing these correlations in $D=2$. In cold atomic gases imaged by quantum gas microscopy, our analysis supports the feasibility of measuring the third order density correlation, from which $\chi_F$ can be reliably extracted in systems with as few as around 100 atoms. For solid-state electron gases, we propose measuring correlations in the speckle pattern of intensity fluctuations in nonlinear X-ray scattering experiments.
\end{abstract}

\maketitle

\section{Introduction} \label{sec:intro}

A central objective of modern condensed matter physics is to identify universal features of correlation and response functions that reveal fundamental properties of quantum phases of matter.   
In recent works, we have established that the topology of the Fermi sea of a Fermi gas is reflected in response and correlation functions.   In particular, it was found that the Euler characteristic, $\chi_F$ of the Fermi sea determines a quantized nonlinear response that in principle can be measured in a weakly interacting two dimensional electron gas (2DEG) \cite{Kane2022a}, or in cold atomic gasses \cite{Yang2022, Zhang2022}.  It was later proposed that the same topological invariant shows up in the linear rectified conductance of a linear Josephson junction proximitizing the 2DEG \cite{Tam2022b, Tam2023}.
In a separate work, we showed that $\chi_F$ is reflected in the equal time correlations of the density of a Fermi gas \cite{Tam2022a}.   This result was introduced as a stepping stone towards the computation of the universal topological multipartite entanglement in a Fermi liquid.  Since equal time density correlations can in principle be measured they are of intrinsic interest, irrespective of the entanglement calculation.   A better understanding of this universal behavior, as well as protocols for experimental verification, are therefore called for.

In this paper, we show that the universal density correlations in a Fermi gas follow from a very simple topological argument that clarifies and generalizes our earlier result.  
We consider the $M$'th order equal time correlation function of a $D$ dimensional non-interacting Fermi gas,
\begin{equation}
s_M({\bf q}_1, ..., {\bf q}_{M-1}) = \int \frac{d^D {\bf q}_M}{(2\pi)^D} \langle \rho_{{\bf q}_1} \rho_{{\bf q}_2} ... \rho_{{\bf q}_M}\rangle_c,
\label{smdef}
\end{equation}
where 
\begin{equation}
\rho_{\bf q} =  \int \frac{d^D{\bf k}}{(2\pi)^D} c^\dagger_{\bf k} c_{{\bf k}+{\bf q}}
\label{rhoqdef}
\end{equation}
and the subscript $c$ indicates the connected correlation function.   Translation symmetry dictates that the correlation function is proportional to $(2\pi)^D \delta({\bf q}_1 + ... + {\bf q}_M)$, so the integral evaluates the coefficient, which is a function of any $M-1$ of the ${\bf q}$'s.

For {\it sufficiently small} ${\bf q}$ (which will be specified below) $s_M$ can be evaluated exactly.
For $M = D+1$,
\begin{equation}
s_{D+1}({\bf q}_1, ..., {\bf q}_D) = \frac{V_{\{\bq_a\}}}{(2\pi)^D} \chi_F,
\label{sd+1}
\end{equation}
where $\chi_F$ is the Euler characteristic of the $D$ dimensional Fermi sea.  $V_{\{\bq_a\}}$ is the volume of the $D$ dimensional parallelepiped formed by $\{\bq_a\}$, given by $\abs{\det\mathbb Q}$, where $\mathbb Q = [\bq_1, ... , \bq_D ]$ is the $D\times D$ matrix of column vectors ${\bf q}_a$.   This relation was established in Ref. \onlinecite{Tam2022a}  for $D=2$ and $D=3$.  In both cases, the calculation was rather elaborate, and for $D=3$ it was only proven in the limit ${\bf q}_a \rightarrow 0$, though numerical results suggested it was exact for finite ${\bf q}_a$.
Here we will provide a much simpler proof of this fact, which can be generalized to all dimensions $D$. The central idea is to \textit{triangulate} the Fermi sea by tiling it with a collection of points, links and triangles (or more generally $d$-dimensional simplexes). We will see that the $D+1$ point density correlation possesses an algebraic structure in terms of the Fermi-Dirac distribution functions that provides precisely a $D$-dimensional triangulation of the Fermi sea, which in turn evaluates a topological invariant in a manner similar to the characterization of polyhedra originally introduced by Euler \cite{Euler1758}.

As a corollary, we will show that for $M > D+1$
\begin{equation}
s_{M}({\bf q}_1, ..., {\bf q}_{M-1}) = 0,
\label{sm>d+1}
\end{equation}
and for $M< D+1$
\begin{equation}
s_{M}({\bf q}_1, ..., {\bf q}_{M-1}) = \frac{V_{\{\bq_a\}}}{(2\pi)^D} \int d^{D+1-M} {\bf k}_\perp \tilde\chi_F({\bf k}_\perp), 
\label{sm<d+1}
\end{equation}
where now the volume $V_{\{\bq_a\}}$ is expressed in terms of the $D\times (M-1)$ matrix  $\mathbb Q = [\bq_1, ... ,\bq_{M-1}]$ as $V_{\{\bq_a\}}=(\det[{\mathbb Q}^T {\mathbb Q}])^{1/2}$.   The integral is taken over the $D+1-M$ dimensional space perpendicular to all of the ${\bf q}_a$.   $\tilde\chi_F({\bf k}_\perp)$ is the Euler characteristic of the intersection of the Fermi sea and the $M-1$ dimensional plane spanned by $\{ {\bf q}_a \}$ that passes through ${\bf k}_\perp$.

In Sec. \ref{sec: Theory} we will consider the physically relevant cases in $D=1,2,3$.   We will begin by establishing Eq. \eqref{sd+1} for $M=D+1$, and then proceed to general $M$.   In Sec. \ref{sec: Exp} we will propose two potential experimental venues for observing these correlations.   The first proposal involves the real space imaging of cold gasses of fermionic atoms, and the second involves measuring the correlations in a fluctuating speckle pattern formed by X-rays scattering from an electron gas.
We will close in Sec. \ref{sec: Discussion} with a brief discussion of the effect of interactions on our results.

\section{Universal Correlations in $D=1,2,3$}\label{sec: Theory}

In this section we outline the proof of the universal density correlation formula.   
We will first focus on the case $M=D+1$, in Eq. \eqref{sd+1} for $D=1, 2$ and $3$.  
A more general proof for $D$ dimensions is presented in Appendix \ref{appendix: general D}.   We will then use this result to establish Eqs. \eqref{sm>d+1} and \eqref{sm<d+1} for $M\ne D+1$.

\subsection{D=1: $s_2(q)$}

In one dimension, it is an elementary exercise to evaluate the equal time density correlation function of the density, $s_2(q)$ for a Fermi gas using Eqs. \eqref{smdef} and \eqref{rhoqdef}.
Using Wick's theorem, the connected correlation function contains a single contraction, given by
\begin{equation}
s_2(q) = \int \frac{dk}{2\pi} (1- f_{k+q}) f_k,
\label{s2int}
\end{equation}
where we use $\langle c^\dagger_\bk c_{\bk'}\rangle = (2\pi)^D \delta(\bk-\bk') f_\bk$, with the Fermi-Dirac distribution function $f_\bk = \theta(E_F- E_\bk)$. This is simply the length in momentum space that is inside the Fermi sea, but outside the Fermi sea when shifted
by $q$. Provided $|q|$ is smaller than the smallest distance between Fermi points, the integral will result in $|q|/(2\pi)$ times the number of disconnected components of the Fermi sea, leading to
\begin{equation}\label{s2chiF}
s_2(q) = \frac{|q|}{2\pi}\chi_F,
\end{equation}
as in Eq. \eqref{sd+1}.

\subsection{D=2: $s_3({\bf q}_1,{\bf q}_2)$}

\begin{figure}
    \centering
    \includegraphics[width=3in]{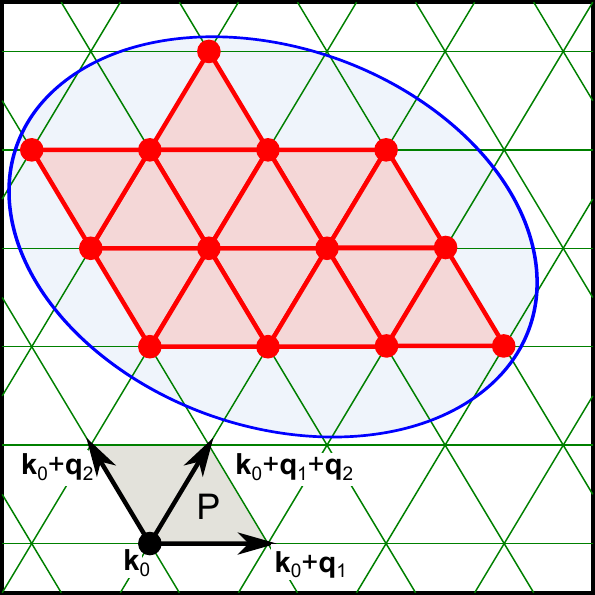}
    \caption{Triangulation of the momentum space and the Fermi sea in two dimensions. The green grid triangulates the entire momentum space as motivated by the algebraic structure of $s_3(\bq_1,\bq_2)$. The Fermi sea is shown in blue, which is triangulated by the collection of all red points, links and triangles lying within. The evaluation of $s_3$ involves integrating $\bk_0$ over the grey parallelogram (P), and for each triangulation fixed by $\bk_0$, counts \#points $-$ \#links $+$ \#triangles inside the Fermi sea.}
    \label{fig:triangulate}
\end{figure}

In two dimensions it is again straightforward to use Wick's theorem to evaluate $s_3({\bf q}_1,{\bf q}_2)$.   There are two contractions, which leads to
\begin{align}
s_3({\bf q}_1,{\bf q}_2) = \int \frac{d^2{\bf k}}{(2\pi)^2} & \Bigl[ (1-f_{{\bf k}+{\bf q}_1 + {\bf q}_2})(1-  f_{{\bf k}+{\bf q}_1}) f_{\bf k} \nonumber \\
&- (1-f_{{\bf k}+{\bf q_1} + {\bf q}_2}) f_{{\bf k}+{\bf q}_2} f_{\bf k} \Bigr].
\label{s3int}
\end{align}
Note that there is a freedom to shift the origin for ${\bf k}$ independently in each of the two terms.   We have fixed that arbitrary choice by requiring that for each value of ${\bf k}$, the integrand depends only on $f_{\bf k}$ evaluated at ${\bf k}$, ${\bf k}+{\bf q}_1$, ${\bf k}+{\bf q}_2$ and ${\bf k}+{\bf q}_1 + {\bf q}_2$, which are the four corners of the elementary parallelogram generated by ${\bf q}_{1,2}$ at ${\bf k}$ as shown in Fig. \ref{fig:triangulate}.   Since these parallelograms tile the plane, it is convenient to write
\begin{equation}
{\bf k}\rightarrow {\bf k}_{\bf n} = {\bf k}_0 + n_1 {\bf q}_1 + n_2{\bf q}_2
\end{equation}
to express the ${\bf k}$ integral as an integral of ${\bf k}_0$ over a single parallelogram, $P$, times a sum on integers $n_{a=1,2}$ describing the lattice of points in momentum space generated by ${\bf q}_{1,2}$
\footnote{\label{note1}Here we assume for simplicity that ${\bf q}$ is a unit fraction of a primitive reciprocal lattice vector ${\bf G}$, so that the lattice generated by ${\bf q}$ fits into the Brillouin zone.  In Appendix \ref{appendix: general D} we show that the argument can be easily generalized to allow  ${\bf q}$ to be any {\it rational fraction} of ${\bf G}$.}.   
Introducing the shorthand notation $f_0 \equiv f_{{\bf k}_{\bf n}}$, $f_a \equiv f_{{\bf k}_{\bf n}+{\bf q}_a}$,
$f_{ab}\equiv f_{{\bf k}_{\bf n}+{\bf q}_a+{\bf q}_b}$, the result can be written as
\begin{equation}
s_3 = \int_{P} \frac{d^2{\bf k}_0}{(2\pi)^2} \sum_{\{n_a\}}
\Bigl[  f_0 - f_0 ( f_1 + f_2 + f_{12}) 
\label{s3sum}
+ f_0 (f_1 + f_2)f_{12}  \Bigr]. 
\end{equation}
Here, we have multiplied out the terms in Eq. \eqref{s3int} and grouped them according to the number of $f$'s.  As seen in Fig. \ref{fig:triangulate}, each parallelogram is associated with one point, three links and two triangles.  Since $f_{\bf k} = 1 (0)$ for ${\bf k}$ inside (outside) the Fermi sea, the three groups of terms altogether count the number of points minus the number of links plus the number of triangles associated with the parallelogram at ${\bf k}_{\bf n}$ that are inside the Fermi sea.

For fixed ${\bf k}_0$, the set of points, links and triangles associated with the lattice ${\bf k}_n$ that are inside the Fermi sea defines a {\it triangulation} of the Fermi sea.    The sum on $n_{a=1,2}$  precisely evaluates Euler's formula - originally introduced to characterize polyhedra \cite{Euler1758, Nakahara1990, Nash1988}, and later generalized -  for the Euler characteristic of the triangulated Fermi sea:
\begin{equation}
\chi_F = \#{\rm points} - \#{\rm links} + \#{\rm triangles}.
\end{equation}
Note that as ${\bf k}_0$ is varied, the number of points, links and triangles in the triangulation will individually change as points pass from inside to outside the Fermi sea.   However, as long as the triangulation faithfully captures the Fermi sea topology, the sum is guaranteed to be unchanged.
Therefore, $\chi_F$ can be factored out of the integral over ${\bf k}_0$, which then simply gives the area of the elementary parallelogram, $|{\bf q}_1 \times {\bf q}_2|$.   We thus conclude that
\begin{equation}
s_3({\bf q}_1,{\bf q}_2)  = \frac{|{\bf q}_1 \times {\bf q}_2|}{(2\pi)^2} \chi_F,
\label{s3chiF}
\end{equation}
as in Eq. \eqref{sd+1}.

As in one dimension, this argument will break down if $|{\bf q}_a|$ is larger than the size of the Fermi sea.   In that case, the triangulation can miss the Fermi sea.   Intuitively, it is clear that for a sufficiently fine mesh, the triangulation will have the correct topology, and Eq. \eqref{s3chiF} will be exact.   Generically, Eq. \eqref{s3chiF} is exact if ${\bf q}_{1,2}$ are smaller than an amount of order $k_F$, the size of the Fermi sea, but precisely how small ${\bf q}_{1,2}$ needs to be depends on the details of the shape of the Fermi sea.   This approach will also break down if ${\bf q}_1 \parallel {\bf q}_2$, since in that case the lattice generated by ${\bf q}_{1,2}$ is degenerate, so Eq. \eqref{s3sum} fails.   In fact, the criterion for how small ${\bf q}_{1,2}$ must be becomes more restrictive if ${\bf q}_1$ and ${\bf q}_2$ are nearly parallel.   In that case, the critical value depends on the curvature of the Fermi surface.   We will study this issue in detail in Appendix \ref{appendix:bound}.

This argument can clearly be implemented in any dimension.  For $D=1$, the integral (\ref{s2int}) can be interpreted as $|q|/(2\pi)$ times the number of points minus the number of links of a triangulation of the one dimensional Fermi sea.   We will present the general proof of this result in Appendix \ref{appendix: general D}, but for clarity we will next outline the case for $D=3$.

\subsection{D=3: $s_4({\bf q}_1,{\bf q}_2,{\bf q}_3)$ }

Our analysis for three dimensions closely follows the preceding section.  
The Wick's theorem expansion of  $s_4(\{{\bf q}_{a=1,2,3}\})$ has 6 terms:
\begin{align}
s_4(\{{\bf q}_{a}\}&) = \int \frac{d^3{\bf k}}{(2\pi)^3} 
\bar f_{{\bf k}+{\bf q}_1 + {\bf q}_2 + {\bf q}_3}  \Bigl[ \nonumber \\
 &\bar f_{{\bf k}+{\bf q}_1 + {\bf q}_2} \bar f_{{\bf k}+{\bf q}_1}  
- f_{{\bf k} + {\bf q}_1 + {\bf q}_3} \bar f_{{\bf k} + {\bf q}_1}  \label{s4int}\\
- &\bar f_{{\bf k} + {\bf q}_1 + {\bf q}_2}  f_{{\bf k} + {\bf q}_2} 
-f_{{\bf k} + {\bf q}_2 + {\bf q}_3} \bar f_{{\bf k} + {\bf q}_2}  \nonumber\\
- &\bar f_{{\bf k} + {\bf q}_1 + {\bf q}_3}  f_{{\bf k} + {\bf q}_3}  
+  f_{{\bf k} + {\bf q}_2 + {\bf q}_3} f_{{\bf k} + {\bf q}_3} 
\Bigr]
f_{{\bf k}}, \nonumber
\end{align}
where we write $\bar f \equiv 1-f$ and we choose the shift in ${\bf k}$ such that each term involves $f_{\bf k}$ evaluated at the corners of the elementary parallelepiped generated by ${\bf q}_{a=1,2,3}$ at ${\bf k}$. 

We next express the integral over ${\bf k}$ as a sum over the lattice generated by ${\bf q}_a$ times an integral of ${\bf k}_0$ over a single unit cell $P$, via 
\begin{equation}
    {\bf k}\rightarrow {\bf k}_{\bf n} = {\bf k}_0 + \sum_{a=1}^3 n_a \bq_a,
\end{equation}
where $n_a \in \mathbb{Z}$. Expanding the terms in Eq. \eqref{s4int} using $\bar f = 1-f$ and using a similar shorthand notation $f_{ab...} \equiv f_{{\bf k}_{\bf n} + {\bf q}_a + {\bf q}_b + ...}$, this gives
\begin{align}
s_4&(\{{\bf q}_a\}) = \int_P \frac{d^3{\bf k}_0}{(2\pi)^3} \sum_{\{n_a\}} f_0  \Bigl\{ 1
\nonumber\\
&-  \Bigl[f_1 + f_2 + f_3 + f_{12} + f_{13} + f_{23}+ f_{123} \Bigr] \label{s4sum}\\
&+  \Bigl[(f_1+f_2)f_{12} + (f_1+f_3)f_{13} + (f_2+f_3)f_{23} \nonumber \\ 
&\quad\quad +(f_1+f_2+f_3+f_{12}+f_{13}+f_{23})f_{123}\Bigr] \nonumber\\
&- \Bigl[(f_1+f_2)f_{12}  + (f_1+f_3)f_{13} + (f_2+f_3)f_{23}\Bigr]f_{123} \Bigr\}. \nonumber
\end{align}
There is 1 term with a single $f$, minus 7 terms with two $f$'s plus 12 terms with three $f$'s minus 6 terms with four $f$'s.   
These correspond precisely to the one point (0-simplex), 7 links (1-simplex), 12 triangles (2-simplex) and 6 tetrahedra (3-simplex) associated with the elementary parallelepiped generated by ${\bf q}_{1,2,3}$ at ${\bf k}$.    The sum over $n_{1,2,3}$ thus evaluates the Euler-Poincar\'e formula for the Euler characteristic
\begin{equation}
\chi_F = \sum_{d=0}^3  (-1)^d \times (\# d{\rm -simplexes})
\end{equation}
associated with the triangulation of the three dimensional Fermi sea based on the lattice generated by ${\bf q}_{1,2,3}$ with origin ${\bf k}_0$.   Assuming ${\bf q}_{1,2,3}$ are small enough such that the triangulation faithfully represents the topology of the Fermi sea, $\chi_F$ is independent of ${\bf k}_0$, and the remaining integral over ${\bf k}_0$ gives the volume $|{\bf q}_1 \cdot ({\bf q}_2 \times {\bf q}_3)|$ of the elementary parallelepiped.  Thus,
\begin{equation}\label{s4chiF}
s_4({\bf q}_1,{\bf q}_2,{\bf q}_3) = \frac{|{\bf q}_1 \cdot ({\bf q}_2 \times {\bf q}_3)|}{(2\pi)^3}\chi_F.
\end{equation}

\subsection{$M>D+1$}

We now consider higher order equal time density correlation functions Eq. \eqref{smdef} with $M>D+1$.   To this end, we consider an auxiliary `decoupled layer' construction by imagining a dimension $\tilde D = M-1$ system consisting of a $\tilde D -D$ dimensional lattice of decoupled $D$ dimensional systems.   The density in this auxiliary system is defined in real space by
\begin{equation}
\tilde \rho \bigl(\tilde{\bf r}= ({\bf r},{\bf r}_\perp)\bigr) \equiv \sum_{\bf m} \rho_{\bf m}({\bf r})\delta^{\tilde D-D}({\bf r}_\perp - {\bf R}_{\bf m}),
\end{equation}
where $\tilde {\bf r}$ is a coordinate in the $\tilde D$ dimensional auxiliary space specified by ${\bf r}$, a coordinate in the original $D$ dimensional space, and ${\bf r}_\perp$, a $\tilde D -D$ dimensional vector defined at lattice points ${\bf R}_{\bf m}$ indexed by integers ${\bf m}=(m_1, ... m_{\tilde D-D})$.   $\rho_{\bf m}({\bf r})$ is the $D$ dimensional density on the ${\bf m}$'th layer.

Since the layers are decoupled, the connected correlations of the density will only be nonzero if all densities are on the same layer.   It follows that density correlations in momentum space, defined as a function of momenta $\tilde {\bf q} = ({\bf q},{\bf q}_\perp)$ are independent of ${\bf q}_\perp$, and are determined by the $D$ dimensional correlations of a single layer.  It follows that
\begin{equation}
\tilde s_{\tilde D+1}(\tilde {\bf q}_1,...,\tilde {\bf q}_{\tilde D}) =\frac{N_\perp}{v_\perp} s_M({\bf q}_1,...,{\bf q}_{M-1}),
\end{equation}
where we used $\tilde D+1 = M$, and ${\bf q}_{1, ...,M-1}$ are $D$ dimensional vectors.   $N_\perp$ and $v_\perp$ are the number of layers and the unit cell volume of the lattice ${\bf R}_{\bf m}$.   $s_M({\bf q}_1,...,{\bf q}_{M-1})$ is the $D$ dimensional correlation function we wish to evaluate.

We can use Eq. \eqref{sd+1} to express $\tilde s_{\tilde D+1}$ in terms of the Euler characteristic $\tilde \chi_F$ of the $\tilde D$ dimensional Fermi sea,
\begin{equation}
\tilde s_{\tilde D+1}(\tilde {\bf q}_1,...,\tilde {\bf q}_{\tilde D+1}) = 
\frac{V_{\{\tilde\bq_i\}}}{(2\pi)^{\tilde D}} \tilde\chi_F.
\end{equation}
However, since the layers are decoupled, the $\tilde D$ dimensional Fermi sea is simply the Cartesian product of the $D$ dimensional Fermi sea and a $\tilde D - D$ dimensional torus.   It follows that $\tilde \chi_F = \chi_F \chi_{T^{\tilde D - D}}$, where the Euler characteristic of the torus is $\chi_{T^{\tilde D - D}} = 0$.   Eq. \eqref{sm>d+1} follows.

\subsection{$M<D+1$}
Finally, we consider a lower order equal time correlation function with $M<D+1$.   In this case, the vectors ${\bf q}_1,...{\bf q}_{M-1}$ span a $\tilde D = M-1$ dimensional subspace of the $D$ dimensional space. The entire momentum space can thus be viewed as a stack of these $\tilde D$ dimensional ``planes", each labeled by a vector $\bk_\perp$ (in its orthogonal complement) that passes through the plane. The Fermi sea can be accordingly viewed as a stack of $ \tilde D$ dimensional cross-sections, each equipped with an Euler characteristic $\tilde \chi_F(\bk_\perp)$ which can be probed by the density correlation $\Tilde{s}_{\Tilde{D}+1}$ restricted to this subspace. It is then straightforward to see that $s_M$ evaluates the \textit{integrated} Euler characteristic along the $\bk_\perp$ directions. 

\begin{figure}
    \centering
    \includegraphics[width=\columnwidth]{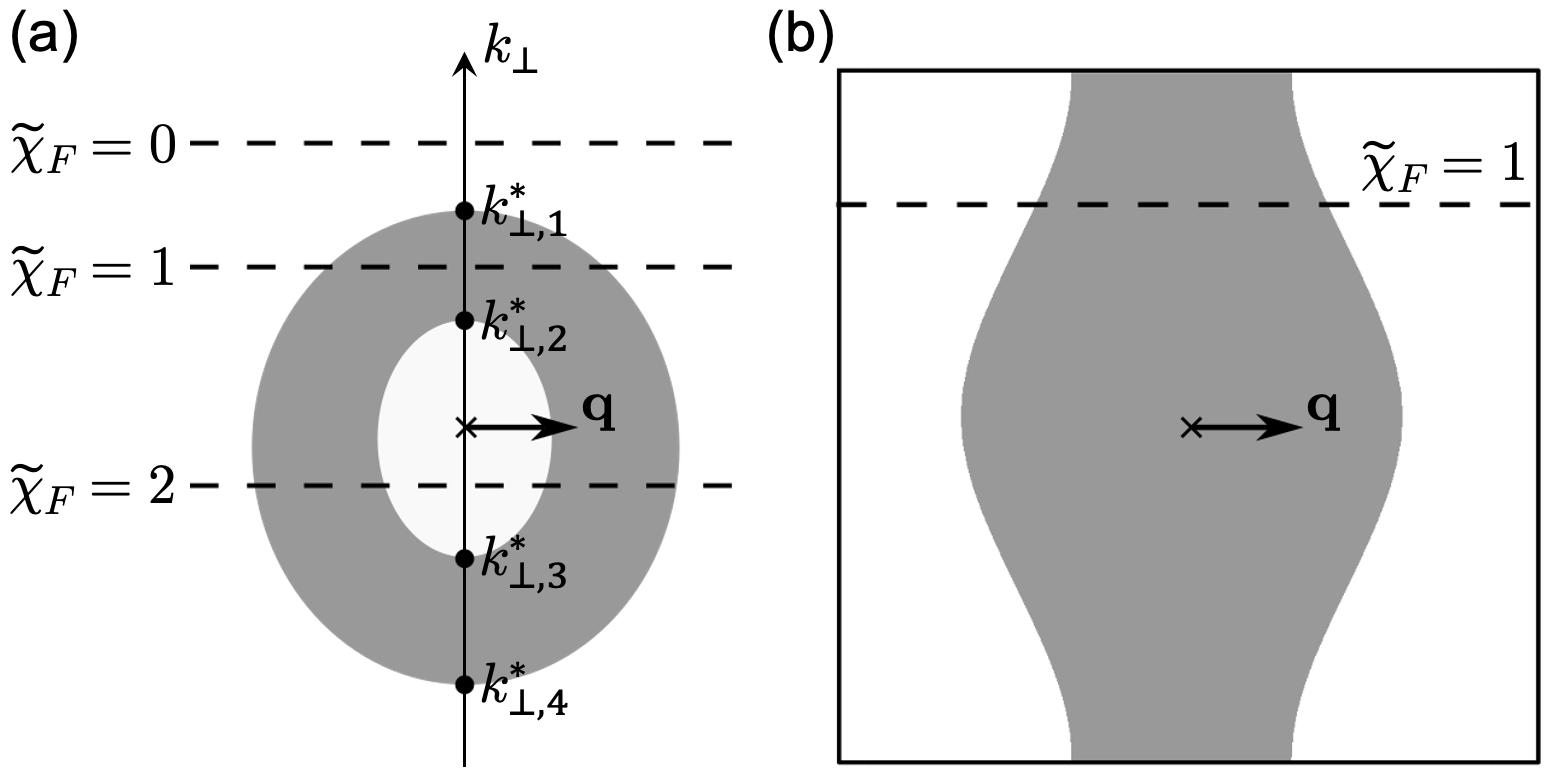}
    \caption{Lower order density correlation probes the integrated Euler characteristic of the lower-dimensional cross-sections of Fermi sea. Two Fermi seas in 2D with $\chi_F=0$ are shown in grey, and the two-point density correlation is being considered. (a) Annulus Fermi sea with closed Fermi surfaces. The Euler characteristic of the cross-section $\tilde \chi_F (k_\perp)$ takes different values, and $s_2(\bq) = (2\pi)^{-2}\abs{\bq}(k^*_{\perp,1}+k^*_{\perp,2}-k^*_{\perp,3}-k^*_{\perp,4})$. (b) Annulus Fermi sea with open Fermi surfaces. Here $\tilde \chi_F =1$ for all $k_\perp$, thus $s_2(\bq) = \abs{\bq}/(2\pi a)$ where $2\pi/a$ is the size of the Brillouin zone (represented as the square frame).}
    \label{fig:sm<d+1}
\end{figure}

Let us consider the $D=M=2$ case for concreteness. Similar to the 1D version in Eq. \eqref{s2int}, for 2D we have
\begin{equation}
    s_2(\bq) = \int \frac{d^2\bk}{(2\pi)^2}  (f_{\bk} - f_{\bk} f_{\bk +\bq}).
\end{equation}
Notice that the structure of the integrand (as a linear combination of products of $f_\bk$) is solely determined by $M$ through Wick's theorem, and is independent of $D$. But now we have an integral over the 2D Brillouin zone, which can be decomposed as $\int d^2\bk \rightarrow \int d \bk_\perp \int d \bk_\parallel$, where $\bk_\perp$ ($\bk_\parallel$) is the 1D vector along the direction perpendicular (parallel) to $\bq$. As such, 
\begin{equation}
    s_2(\bq) = \int \frac{d \bk_\perp}{2\pi} \tilde s_2(\bq; \bk_\perp) = \frac{\abs{\bq}}{(2\pi)^2} \int d \bk_\perp \Tilde{\chi}_F(\bk_\perp), 
    \label{s2d=2}
\end{equation}
where $\tilde s_2(\bq; \bk_\perp)$ is interpreted as a 2-point density correlation restricted to the 1D cut through the momentum space at $\bk_\perp$, and is evaluated by Eq. \eqref{s2chiF}. Figure \ref{fig:sm<d+1} illustrates the application of this formula to probe the combined geometric and topological structure of 2D Fermi seas. 

For the general case with $M < D+1$, we can express
\begin{equation}
s_M({\bf q}_1,...,{\bf q}_{M-1}) = \int \frac{d^{D-\tilde D} {\bf k}_\perp}{(2\pi)^{D-\tilde D}} 
\tilde s_{\tilde D+1}({\bf q}_1,...,{\bf q}_{\tilde D}; {\bf k}_\perp), 
\end{equation}
where $\tilde s_{\tilde D+1}$ is the $(\tilde D+1)$-point density correlation restricted to the $\tilde D$-dimensional subspace at $\bk_\perp$. We then apply Eq. \eqref{sd+1} (with $D \mapsto \Tilde{D}$) to evaluate $\tilde s_{\tilde D+1}$ and arrive at Eq. \eqref{sm<d+1}.

Before we conclude the formal discussion on the structure of universal density correlations, let us remark on the validity of Eq. \eqref{sm<d+1}: for generic Fermi seas, the equality is exact only in the limit $\bq \rightarrow 0$. This should be distinguished from the broader range of validity of Eq. \eqref{sd+1} (for $M=D+1$) and Eq. \eqref{sm>d+1} (for $M>D+1$), which hold for finite $\bq$'s (whose bounds are determined by the precise shape of the Fermi surface, see App. \ref{appendix:bound}). The reason is that, when integrating over $\bk_\perp$, the quantized value of $\Tilde{\chi}_F (\bk_\perp)$ can change at some $\bk^*_\perp$ due to the presence of closed Fermi surfaces, as shown in Fig. \ref{fig:sm<d+1}(a). When $\bk_\perp$ approaches $\bk^*_\perp$, the $\tilde D$-dimensional cross-section of the Fermi sea would become so tiny (or different components can get so close) that the application of Eq. \eqref{sd+1} on this subspace is no longer exact for any finite $\bq$. These regions gives rise to corrections higher order in $\bq$ beyond $V_{\{\bq_i\}} \sim \mathcal{O}(q^M)$. Nevertheless, for certain situations with only open Fermi surfaces, Eq. \eqref{sm<d+1} can still hold exactly for finite $\bq$, as illustrated in Fig. \ref{fig:sm<d+1}(b).

\section{Proposed Experiments}\label{sec: Exp}
Next we suggest two experimental avenues for probing Fermi sea topology based on the aforementioned multi-point density correlation. The first proposal applies to Fermi gases realized in optical lattices and relies on the advancement of quantum gas microscopy \cite{Bakr_review}, which allows for imaging of a single fermionic atom in a degenerate Fermi gas. The second proposal is applicable to solid-state systems and makes use of non-linear scattering of X-rays, which probes the topological density correlation of electrons from the fluctuating ``speckle pattern'' of the scattered waves.

\subsection{Imaging Cold Atomic Gasses}
Arguably the most straightforward way to measure the density correlation is by \textit{counting} the particle number. While counting individual electrons in a 2DEG remains unfeasible, site-resolved fluorescence imaging of individual fermionic atoms (such as $\prescript{6}{}{}\text{Li}$ and $\prescript{40}{}{}\text{K}$) in a many-body setting including several hundreds to thousands of atoms has been achieved since the pioneering experimental works in 2015 \cite{Cheuk2015, Haller2015, Parsons2015, Edge2015}. This technique, known as the quantum gas microscopy, has enabled direct observation of Mott insulators \cite{Greif2016, Cheuk2016a}, as well as measurements of the spatial spin and charge correlations \cite{Cheuk2016b, Parson2016}, in the Fermi-Hubbard model. In a spin-polarized degenerate Fermi gas, Pauli blocking has also been established by observing strongly suppressed on-site atom number fluctuations \cite{Omran2015}. With this atomic resolution, here we propose to first extract the connected 3-point real-space density correlation from images of a 2D ultracold Fermi gas,
\begin{equation}
\begin{split}
\mathfrak{s}_3(\br_1,\br_2,\br_3) = & \ \langle \rho_{\br_1}\rho_{\br_2} \rho_{\br_3}\rangle_c \\
= & \ \langle \rho_{\br_1}\rho_{\br_2} \rho_{\br_3}\rangle 
- \langle \rho_{\br_1}\rho_{\br_2}\rangle\langle \rho_{\br_3}\rangle \\
&- \langle \rho_{\br_2}\rho_{\br_3}\rangle\langle \rho_{\br_1}\rangle 
- \langle \rho_{\br_1}\rho_{\br_3}\rangle\langle \rho_{\br_2}\rangle \\
&+ 2\langle \rho_{\br_1}\rangle\langle\rho_{\br_2}\rangle\langle \rho_{\br_3}\rangle.
\end{split}
\end{equation}
By Fourier transformation, the momentum-space density correlation can then be obtained, with the momentum resolution controlled by the inverse system size $L^{-1}$. The exact small-$q$ behavior predicted in Eq. \eqref{s3chiF} should become observable for large enough systems with $k_F \gg L^{-1}$.

To demonstrate the feasibility of this proposal, we present below simulations of a tight-binding model on a square lattice with nearest-neighbour hoppings and open boundary conditions. With lattice size $L \times L$, we focus on the $L=14$ case for concreteness, though the predicted momentum dependence of the form $\chi_F \abs{\bq_1 \times \bq_2}$ can be observed even in smaller systems, such as $L=10$. Finite-temperature effects at the order of $k_BT \sim 0.1 E_F$ are also considered, showing that major \textit{quantitative} features of the topological density correlation remains observable in realistic settings. 

\begin{figure}
    \centering
    \includegraphics[width=\columnwidth]{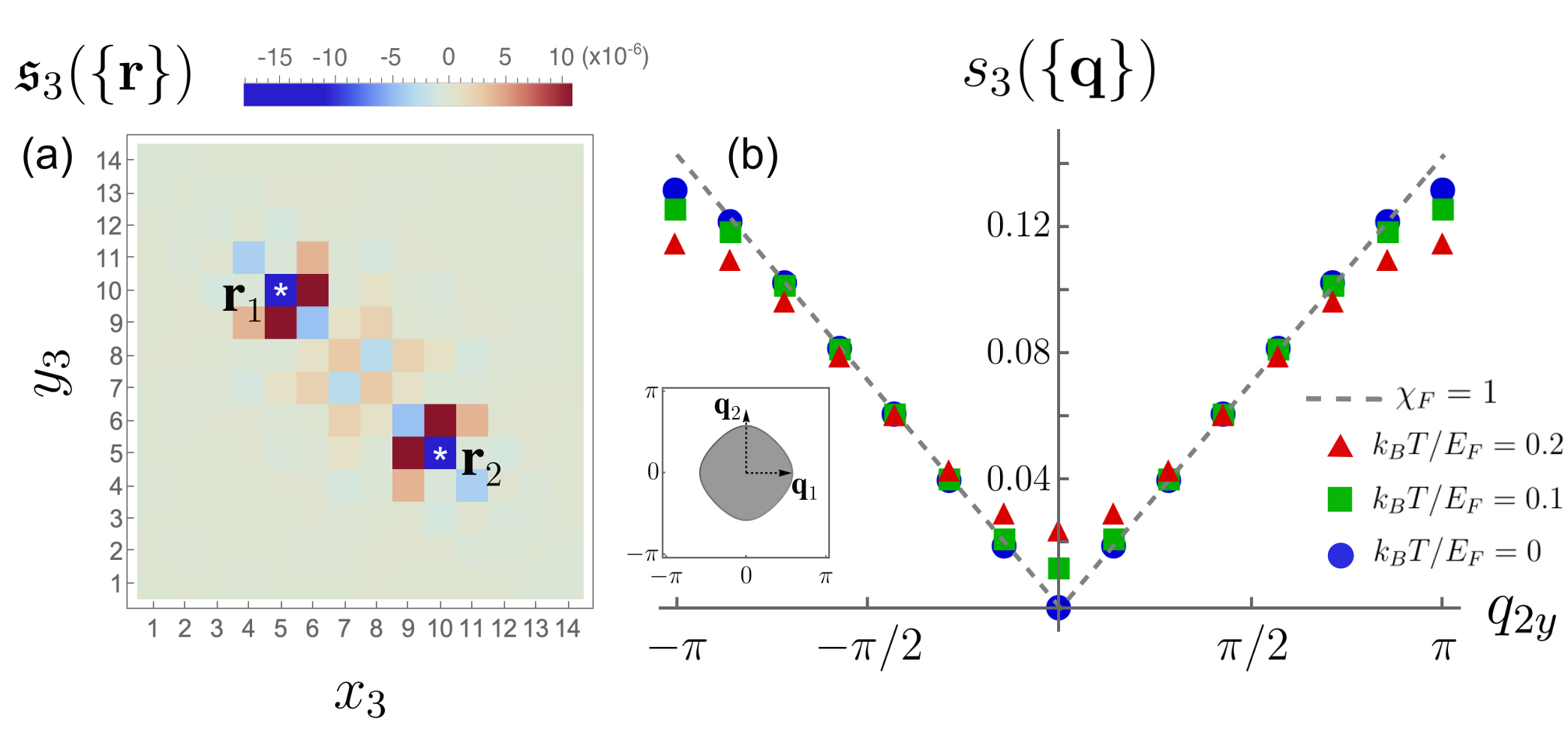}
    \caption{Simulation of the quantum gas microscopy measurement for the topological density correlation in a square-lattice tight-binding model with nearest neighbour hopping $-t$ and size $14\times 14$. The quantum gas is treated as a grand canonical ensemble at chemical potential $\mu=-4t+E_F$, with $E_F=2.5t$. (a) Real-space 3-point density correlation $\mathfrak{s}_3(\br_1,\br_2,\br_3)$ versus $\br_3$, with $\br_1=(5,10)$ and $\br_2=(10,5)$, simulated at temperature $k_BT= 0.1E_F$. (b) Momentum-space density correlation $s_3(\bq_1,\bq_2,-\bq_1-\bq_2)$, see Eq. \eqref{eq:dFFT}, as a function of $q_{2y}$ by taking $\bq_1=(4\pi/7,0)$ and $\bq_2=(0,q_{2y})$, computed at various temperature. Dashed line corresponds to the zero-temperature infinite-volume prediction in Eq. \eqref{s3chiF}. Inset shows the corresponding Fermi sea with $\chi_F=1$.}
    \label{fig:coldatom_temp}
\end{figure}

The quantity $\mathfrak{s}_3(\{\br\})$, which is observable in a quantum gas microscope, can be numerically computed in this free fermion lattice model by using Wick's theorem to express it in terms of the 
two-point correlator $g_{\br',\br} = \langle c^\dagger_{\br'}c_{\br}\rangle$:
\begin{equation}
\begin{split}
    \mathfrak{s}_3(\br_1,\br_2,\br_3) =  \ &
    g_{\br_1,\br_2}g_{\br_2,\br_3}g_{\br_3,\br_1}+g_{\br_1,\br_3}g_{\br_3,\br_2}g_{\br_2,\br_1}\\
   & -\delta_{\br_1,\br_2}|g_{\br_2,\br_3}|^2 -\delta_{\br_2,\br_3}|g_{\br_3,\br_1}|^2 \\
   & -\delta_{\br_3,\br_1}|g_{\br_1,\br_2}|^2 
    + \delta_{\br_1,\br_2} \delta_{\br_2,\br_3} g_{\br_1,\br_1}.
\end{split}
\end{equation}
Here we have used $\rho_\br = c^\dagger_\br c_\br$, where $c^\dagger_\br$ creates a fermion at the lattice site $\br$ and the lattice constant has been set to unity.
Treating the cold atomic gas as a grand canonical ensemble at chemical potential $\mu$ and temperature $T$, we have
\begin{equation}
    g_{\br',\br} = \sum_n U^\dagger_{\br',n} f_n U_{n,\br},
\end{equation}
where $f_n = [e^{(E_n-\mu)/(k_BT)}+1]^{-1}$ is the finite-temperature Fermi distribution function for the $n$-th energy-eigenstate, whose normalized eigenvector obtained by diagonalizing the real-space tight-binding Hamiltonian is denoted as $U_{n,\br}$. To illustrate the readily observable quantity from atomic gas imaging, we have plotted $\mathfrak{s}_3(\{\br\})$ as a function of $\br_3$ in Fig. \ref{fig:coldatom_temp}(a), by fixing $\br_1=(5,10)$ and $\br_2 = (10,5)$, for an isotropic tight-binding model ($L=14$) with nearest-neighbor hopping $-t$ ($t>0$) and Fermi energy $E_F=2t$ (as counted from the band-bottom, hence $\mu=-4t+E_F$). This corresponds to an electronlike Fermi surface with $\chi_F=1$. 

For an infinite system, the continuous Fourier transform of Eq. \eqref{s3chiF} has been worked out in Ref. \cite{Tam2022a} as
\begin{equation}
    \mathfrak{s}_3 = \frac{\chi_F}{8\pi^4} \delta''(A_{123}),
    \label{delta''}
\end{equation}
where $A_{123}$ is the area of the triangle formed by $\{\br_1,\br_2, \br_3\}$. The derivative of the delta function suggests that the real-space density correlation is dominated by straight-line configurations in which $\br_1,\br_2$ and $\br_3$ are collinear. Such a qualitative feature is marginally observable in Fig. \ref{fig:coldatom_temp}(a), where $\mathfrak{s}_3$ peaks around the diagonal. 

Though finite-size effects make the detailed comparison of Eq. \eqref{delta''} with the real-space data difficult, the topological nature of the density correlations present in the multi-dimensional data set $\mathfrak{s}_3(\br_1,\br_2,\br_3)$ is much more apparent in the momentum space. The momentum-space density correlation can be computed by the \textit{discrete} Fourier transform,
\begin{equation}\label{eq:dFFT}
    s_3(\bq_1,\bq_2,\bq_3) = \frac{1}{L^2}\sum_{\br_{1,2,3}} \mathfrak{s}_3(\{\br\})e^{-i[\bq_1\cdot \br_1 +\bq_2\cdot \br_2 +\bq_3\cdot \br_3]},
\end{equation}
where ${\bf q} = 2\pi (n_x,n_y)/L$ for integers $n_{x,y}$. Notice that $s_3$ defined here is slightly different from that defined previously for an infinite system. It is not a delta function of $\bq_1+\bq_2+\bq_3$ as the translation symmetry is violated by the open boundary conditions. Nonetheless, we shall focus on $\bq_3=-\bq_1-\bq_2$ where we can clearly see the topological signature predicted for the infinite system. In Fig. \ref{fig:coldatom_temp}(b), with $\bq_1 = (4\pi/7,0)$ and $\bq_2 = (0,q_{2y})$, we have plotted $s_3(\bq_1,\bq_2,-\bq_1-\bq_2)$ as a function of $q_{2y}$ in both the zero-temperature limit (which has been assumed in all the preceding sections) and for experimentally feasible temperatures at fractions of the Fermi energy. As clearly shown in Fig. \ref{fig:coldatom_temp}(b), the singular momentum-dependence $\abs{\bq_1\times \bq_2}$ is observable for a reasonably small system at a finite temperature, from which $\chi_F$ can be reliably extracted through Eq. \eqref{s3chiF}.

\begin{figure}
    \centering
    \includegraphics[width=\columnwidth]{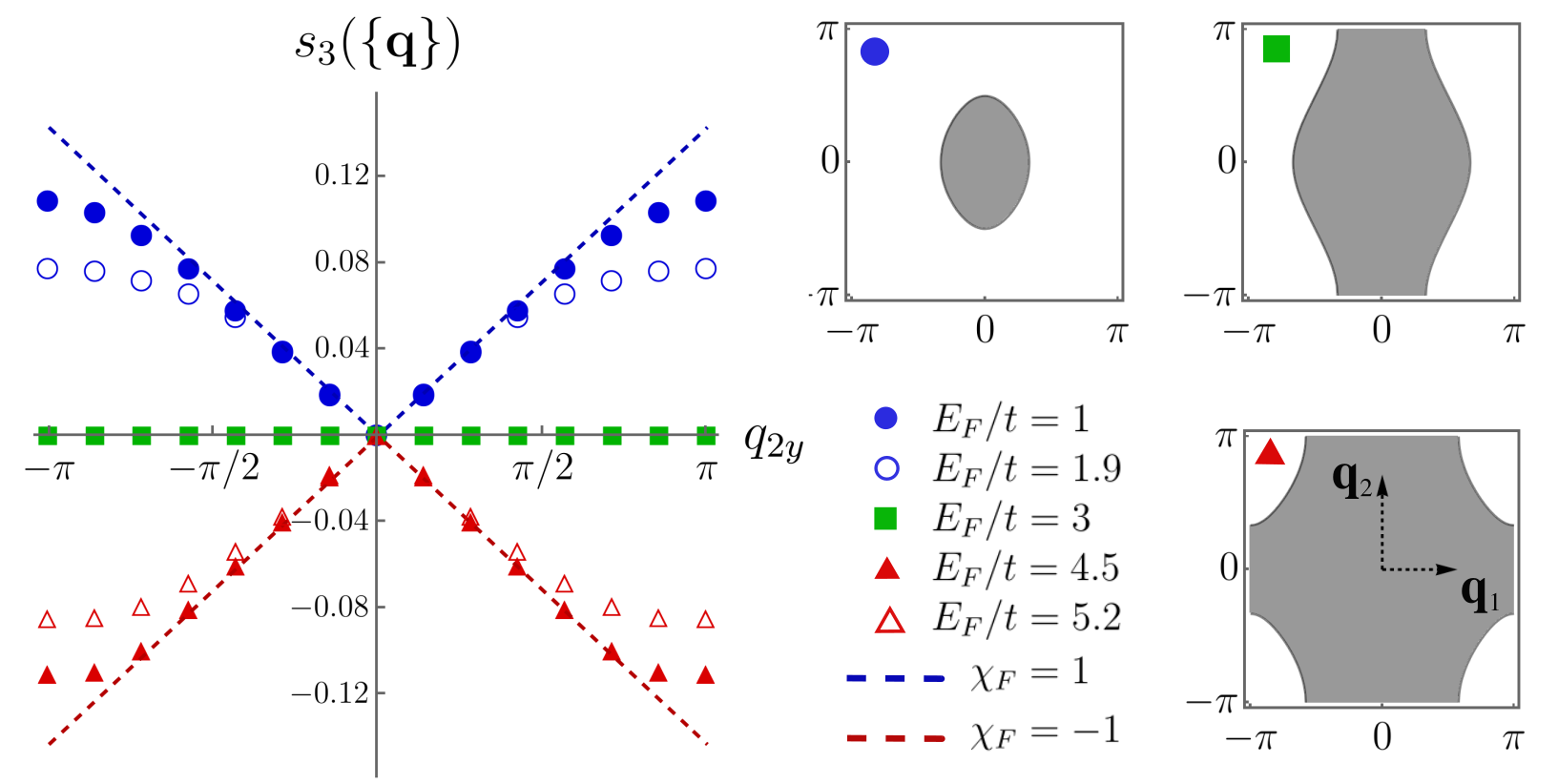}
    \caption{Probing Lifshitz transitions by $s_3(\bq_1,\bq_2,-\bq_1-\bq_2)$ in a $14\times 14$ tight-binding model with anisotropic nearest neighbour hoppings, $t_x=2t_y=t$, and open boundaries. The quantum gas is simulated as a grand canonical ensemble at $\mu=-3t+E_F$ and $T=0$, where the Fermi seas for three simulated $E_F$ are depicted on the right. We have chosen $\bq_1=(4\pi/7,0)$ and $\bq_2=(0,q_{2y})$ for plotting $s_3(\{\bq\})$, and the dashed lines are predictions from Eq. \eqref{s3chiF}.}
    \label{fig:coldatom_transition}
\end{figure}

To further demonstrate the ability for $s_3(\bq_1,\bq_2,-\bq_1-\bq_2)$ to capture the Fermi sea topology in finite systems, we have also considered anisotropic nearest neighbour hoppings, $t_x=2t_y=t$, and computed $s_3(\{\bq\})$ for a set of chemical potentials that realize $\chi_F=\pm1, 0$, see Fig. \ref{fig:coldatom_transition}. In this model, the Fermi sea is topologically equivalent to a disk ($\chi_F=1$) for $2t> E_F > 0$, an annulus ($\chi_F=0$) for $4t> E_F > 2t$, and a punctured torus ($\chi_F=-1$) for $6t>E_F>4t$. It can be seen that the singularity around $q_{2y}=0$ only concerns $\chi_F$, but not the precise geometry of the Fermi sea. The shape of Fermi sea, however, controls the range of validity for Eq. \eqref{s3chiF}, which we discuss in detail in Appendix \ref{appendix:bound}. Particularly, as shown in Fig. \ref{fig:coldatom_transition}, a Fermi sea closer to a Lifshitz transition (cf. open circles) shows $\abs{\bq_1\times \bq_2}$ dependence in a narrower range as compared to a Fermi sea further away from the transition (cf. filled circles). Our analysis thus provides guidance for experimentally demonstrating Eq. \eqref{s3chiF} in a small finite-size system. 

There is another interesting yet more challenging approach to probe $\chi_F$, which requires performing the quantum gas microscopy for a wide range of system sizes. As pointed out in Ref. \cite{Tam2022a}, the momentum-space density correlation in Eq. \eqref{sd+1} implies a topological multipartite \textit{number} correlation, which is in turn related to the multipartite entanglement present in a Fermi gas. For $D=2$, by spatially partitioning the Fermi gas into three mutually neighboring regions ($A, B$ and $C$) that meet at a point, we found the following finite-size scaling:
\begin{equation}
    \langle Q_A Q_B Q_C \rangle_c = \frac{3\chi_F}{8\pi^4} \log^2 \frac{L}{a} + b,
\end{equation}
where $Q_{K=A,B,C}$ is the particle number operator in region $K$, while $a$ and $b$ are non-universal fitting parameters. Importantly, the coefficient of the logarithmic scaling is solely determined by the Fermi sea topology and insensitive to the partition geometry in the real space.

\subsection{Non-Linear X-Ray Scattering} 

\begin{figure}
    \centering
    \includegraphics[width=\columnwidth]{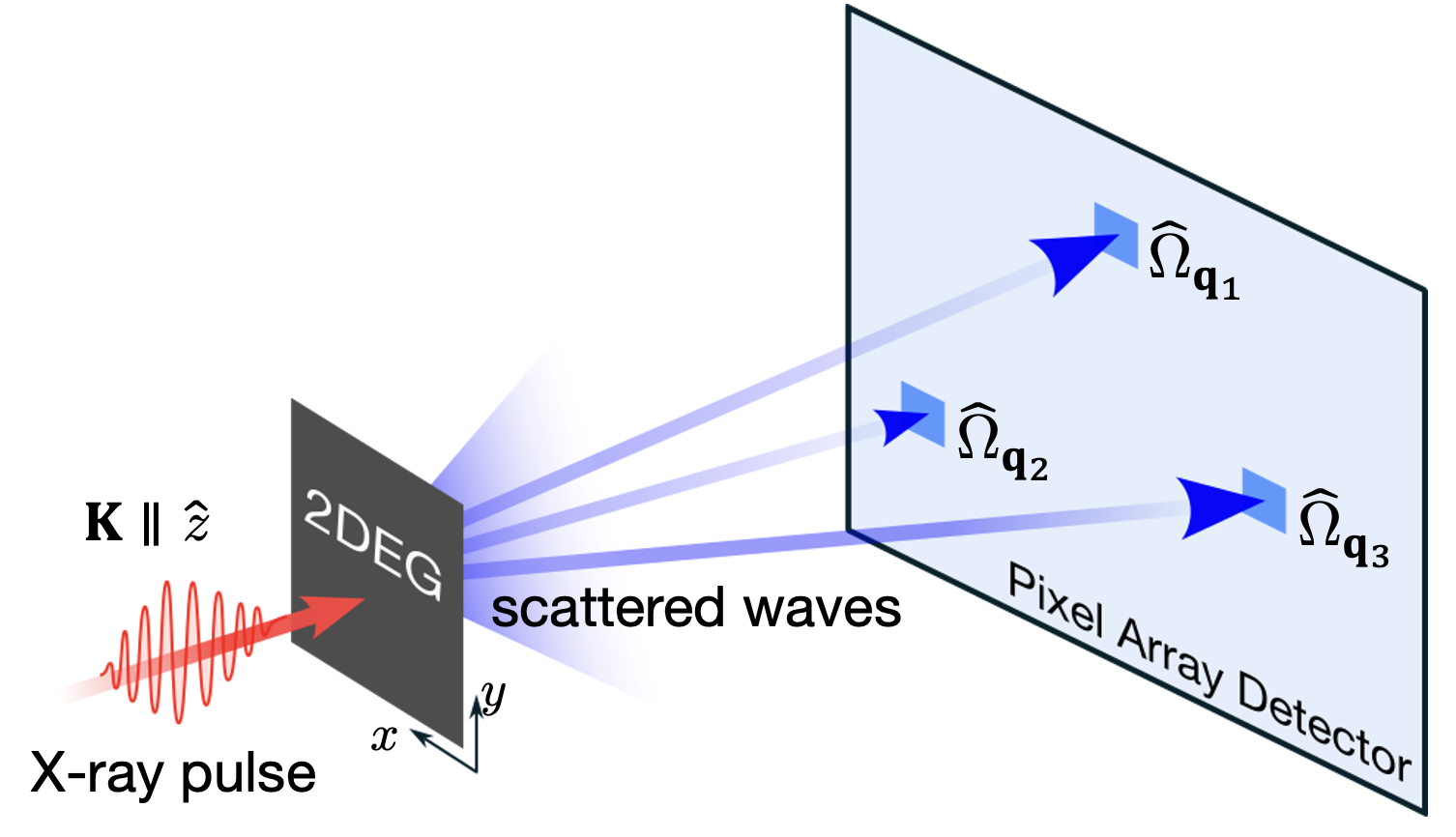}
    \caption{Schematic of nonlinear X-ray scattering for probing topology in a 2DEG. The speckle pattern in the scattered intensity contains fluctuations that encode multi-point electronic density correlations. Whenever $\bq_1 \pm \bq_2 \pm\bq_3 \sim 0$, there is an enhanced third order correlation in $\langle \Delta N_{\bq_1}\Delta N_{\bq_2}\Delta N_{\bq_3} \rangle \propto \chi_F^2$ , where $N_{\bq_i}$ measures the photon number per solid angle scattered into the direction $\hat{\Omega}_{\bq_i}$. }
    \label{fig:Xray}
\end{figure}

We next propose a method for probing the universal density correlations in an electron gas in a solid state system using scattering.   Scattering has long been established as a powerful tool for studying the structure of matter.   By shining high frequency light on matter, each electron becomes an oscillating dipole that radiates on its own, so the spatial correlation among all electrons is imprinted on the interference of the scattered light.   We will argue that X-rays scattered from the electrons in a degenerate Fermi gas contain correlations that reflect the universal density correlations in Eq. \eqref{sd+1}.   For concreteness, we consider a geometry shown in Fig. \ref{fig:Xray}, in which electrons in a two dimensional electron gas (2DEG) scatter X-rays.

The dominant process for the scattering of high energy photons from electrons can be understood by considering the $|\bA|^2$ term in the expansion of the kinetic energy $|\bp-e\bA/c|^2/(2m_e)$.   This results in an interaction term of the form
\begin{equation}
\Delta H = \frac{e^2}{2m_e c^2} \int d\br \rho(\br)|\bA(\br)|^2,
\label{DeltaHA2}
\end{equation}
where $\rho(\br)$ is the electron density \cite{girvin_yang_2019}.
For an incident beam of photons with momentum $\bK$, the rate for photons to scatter to momentum $\bK'=\bK+\bq$ can be computed using Fermi's golden rule.   Integrating over the energies of the scattered photons, the number of photons per unit time and solid angle $dN_\bq/dt$ scattered in a direction $\hat\Omega_\bq = \bK'/|\bK'|$ is
\begin{equation}
    \frac{dN_\bq}{dt} = \frac{I_p}{\hbar \omega}\Big(\frac{e^2}{m_e c^2}\Big)^2 \langle \rho_{-\bq}\rho_\bq\rangle,
    \label{GoldenRule}
\end{equation}
where $I_p$ is the intensity of the incident beam.   When scattering from a free electron gas, this describes an incoherent process, in which a momentum $\bq$ is transfered to an electron.   For large $\bq$, this results in the Compton effect, whereby the scattered photon is red shifted.   For a degenerate Fermi gas, when $|\bq| \ll k_F$, the Pauli principle restricts the transitions to electrons at momentum $\bk$ that cross the Fermi surface when scattered to $\bk - \bq$.  This leads to a suppression of the scattered intensity for $|\bq| < k_F$.   The frequency integrated scattering rate probes precisely the structure factor
\begin{equation}
\langle \rho_{-\bq}\rho_\bq\rangle = L^2 s_2(\bq) \sim \frac{L^2}{2\pi^2}  k_F |\bq|,
\end{equation}
where $L^2$ is the area of the 2DEG, and we have used Eq. \eqref{s2d=2} and considered for simplicity a circular Fermi surface with diameter $2k_F$.

We will argue that the scattered photons for $|\bq| \lesssim k_F$ contain fluctuations with correlations that encode the topological density correlations in the Fermi gas.  This will lead to a kind of {\it speckle pattern} in the scattered intensity that fluctuates both as a function of scattering angle and as a function of time.   Thus, it will be important to be able to resolve the scattering intensity with both angular and temporal resolution.   It appears to us that this is on the boundary of feasibility.   For a 2DEG with $k_F \sim (10\; {\rm nm})^{-1}$, we should resolve $|\bq| \sim (100\; {\rm nm})^{-1}$.   For a soft X-ray with $\hbar\omega = 1 \;{\rm keV}$, $|\bq|/|\bK| \sim 2 \times 10^{-3}$, corresponding to an angle $\theta \sim 0.1^\circ$, which is in range for small angle X-ray scattering \cite{brumberger2013modern, jeffries2021small}.  While going to photons with lower energy and longer wavelength could improve the $\bq$ resolution, there is a tension:  At lower energy virtual (or resonant) processes can become operative and complicate the analysis.   These arise due to coupling to excited states via the first order interaction $e \bp\cdot \bA /(m_e c)$, which was not included in Eqs. \eqref{DeltaHA2} and \eqref{GoldenRule}.  For a given material system it will be desirable to optimize the photon energy.

The time scale for fluctuations of the intensity at $\bq$ will be of order $\tau \sim 1/(v_F |\bq|)$.  For a Fermi velocity of order $v_F \sim 10^6 {\rm m/s}$, the temporal scale is $\tau \sim 100\;{\rm fs}$.   While detecting photons with this temporal resolution seems challenging, pulsed X-rays with pulse widths of femtoseconds are possible \cite{behrens2014few, helml2014measuring, hoffmann2018femtosecond, Pandey_2020}.   We therefore propose to use femtosecond X-ray pulses to effectively take a snapshot of the electrons, and to measure the correlations in the number of scattered photons.  In the following we will develop a theory of those correlations.

Consider a single pulse $I_p(t)$ at time $t_0$, with an integrated energy per unit area, $F_p = \int dt I_p(t)$ and a width $\Delta t$ satisfying $\omega^{-1} \ll \Delta t \ll \tau$.   According to Eq. \eqref{GoldenRule}, the average number of scattered photons per solid angle scattered to $\hat\Omega_\bq$ is
\begin{equation}
\langle N_{\bf q} \rangle = C_0 \langle \rho_{-\bq}\rho_{\bq} \rangle,
\end{equation}
with $C_0 \equiv (F_p/\hbar\omega)(e^2/m_e c^2)^2 $.
The second order correlation in the number of photons detected at $\bq_1$ and $\bq_2$ can be computed similarly:
\begin{equation}
    \langle N_{\bq_1}N_{\bq_2}\rangle = C_0^2
    \langle \rho_{-\bq_1}\rho_{\bq_1} \rho_{-\bq_2}\rho_{\bq_2}\rangle.
    \label{nq2}
\end{equation}
Since we have assumed the pulse width $\Delta t$ is smaller than the characteristic time $\tau$ for $\rho_\bq$ to fluctuate, all of the densities in (\ref{nq2}) are evaluated at the same time $t_0$.   This correlator can be separated into connected and disconnected pieces:
\begin{equation}
\begin{split}
    \langle &\rho_{-\bq_1}\rho_{\bq_1} \rho_{-\bq_2}\rho_{\bq_2}\rangle = 
     \langle \rho_{-\bq_1}\rho_{\bq_1}\rangle_c \langle \rho_{-\bq_2}\rho_{\bq_2}\rangle_c \\
    & + \langle \rho_{-\bq_1}\rho_{\bq_2}\rangle_c\langle \rho_{-\bq_2}\rho_{\bq_1}\rangle_c
     + \langle \rho_{-\bq_1}\rho_{-\bq_2}\rangle_c\langle \rho_{\bq_1}\rho_{\bq_2}\rangle_c \\
     & + \langle \rho_{-\bq_1}\rho_{\bq_1}\rho_{-\bq_2}\rho_{\bq_2}\rangle_c, \\
     \end{split}
     \label{rho4}
\end{equation}
where we have assumed $\langle\rho_{\pm\bq_1}\rangle=\langle\rho_{\pm\bq_2}\rangle=0$.   According to Eq. \eqref{sm>d+1},
the connected four point correlator will vanish for $D=2$.   Moreover, $\langle \rho_{-\bq_1}\rho_{\bq_2}\rangle_c = (2\pi)^2\delta_{\bq_1-\bq_2} s_2(\bq_1)$.   Therefore, we have
\begin{equation}
\begin{split}
      \langle &\Delta N_{\bq_1}\Delta N_{\bq_2}\rangle \equiv  \langle N_{\bq_1}N_{\bq_2}\rangle-\langle N_{\bq_1}\rangle\langle N_{\bq_2}\rangle  \\
     &= (2\pi)^4 C_0^2(\delta^2_{\bq_1-\bq_2}  + \delta^2_{\bq_1+\bq_2} )|s_2(\bq_1)|^2.
      \end{split}
      \label{DeltaN2}
\end{equation}
The delta functions have a width and height that is set by the size $L$ of the 2DEG, so $\delta^2_{\bq}$ has a peak height  $\sim L^4/(2\pi)^4$ and a width $\sim 2\pi/L$.   The term proportional to $\delta_{\bq_1-\bq_2}$ thus predicts that $\langle (\Delta N_\bq)^2 \rangle \sim \langle N_\bq \rangle^2$, with correlations over a range $2\pi/L$ in $q$.   This leads to a speckle pattern in $N_\bq$.   If the pixel resolution of the detectors can resolve $\Delta q \sim 2\pi/L$, then the fluctuation $\langle(\Delta N_\bq)^2\rangle$ is of order $\langle N_\bq\rangle^2$.  If the resolution is lower, then averaging will reduce the size of the observed fluctuations.    In addition to the fluctuations $\langle (\Delta N_\bq)^2 \rangle$, Eq. \eqref{DeltaN2} predicts that the speckle pattern will exhibit enhanced correlations between different scattering directions specified by $\bq_1$ and $\bq_2$ when $\bq_1+\bq_2\sim 0$.   These correlations also probe the second order equal time density correlations $s_2(\bq)$, given in Eq. \eqref{sm<d+1}.  While they do not directly probe the topology of the Fermi sea, observing them would be a prerequisite for observing the topological density correlations.

In $D=2$, the topological density correlations are encoded in the third order equal time density correlation.   This motivates us to consider the third order correlations in the measured intensity, 
\begin{equation}
    \langle N_{\bq_1} N_{\bq_2} N_{\bq_3}\rangle =C_0^3
    \langle \prod_{a=1}^3\rho_{-\bq_a}\rho_{\bq_a} \rangle.
\end{equation}
Again, this will contain connected and disconnected components.   As in Eq. \eqref{rho4}, the 4th and 6th order connected correlations of $\rho_\bq$ will vanish.   If we assume that $\pm\bq_1$, $\pm\bq_2$ and $\pm\bq_3$ are all distinct, then the only non-zero connected pieces involve
\begin{equation}
\begin{split}
     \langle \prod_{a=1}^3&\rho_{-\bq_a}\rho_{\bq_a} \rangle = 
     \langle \rho_{-\bq_1}\rho_{\bq_1} \rangle_c \langle \rho_{-\bq_2}\rho_{\bq_2} \rangle_c \langle \rho_{-\bq_3}\rho_{\bq_3} \rangle_c
      \\
    + &\sum_{\nu_2,\nu_3=\pm 1}\langle \rho_{-\bq_1}\rho_{-\nu_2\bq_2}\rho_{-\nu_3\bq_3}\rangle_c  \langle \rho_{\bq_1}\rho_{\nu_2 \bq_2}\rho_{\nu_3 \bq_3}\rangle_c.
     \end{split}
\end{equation}
Therefore, we predict enhanced third order correlations when $\bq_1\pm \bq_2 \pm \bq_3 \sim 0$:
\begin{equation}
\begin{split}
    \langle \Delta &N_{\bq_1}\Delta N_{\bq_2} \Delta N_{\bq_3}\rangle = \\
    &(2\pi)^4 C_0^3 \sum_{\nu_2,\nu_3=\pm 1} \delta^2_{\bq_1+\nu_2\bq_2+\nu_3\bq_3}|s_3(\bq_1,\nu_2\bq_2)|^2.
    \end{split}
    \label{DeltaN3}
\end{equation}
    While achieving the necessary resolution to resolve these third order correlations presents a technical challenge, Eq. \eqref{DeltaN3} shows that they directly probe the third order equal time density correlations (\ref{s3chiF}), which encode the topological structure of the Fermi gas.

\section{Discussion} \label{sec: Discussion}

In this paper we have studied the structure of the equal time density correlations in a Fermi gas, and presented a simple analysis that explains how they probe the topology of the Fermi sea.   For a Fermi gas in $D$ dimensions, the $D+1$ point density correlation expressed 
in momentum space 
is universal, and is exactly quantized for sufficiently small momenta $\{\bq_a\}$, with a coefficient given by the Euler characteristic of the Fermi sea.   
 For $M<D+1$ the $M$ point correlation is related to the topological correlations in $\tilde{D}= M-1$ dimensional slices of the Fermi sea parallel to $\{\bq_a\}$, which leads to a result that scales with the $D-\tilde{D}$ dimensional volume of the Fermi sea perpendicular to $\{\bq_a\}$.   For $M>D+1$, we found that the $M$ point correlation vanishes.

The vanishing of the density correlations when $M>D+1$ is reminiscent of the  method of bosonization for $D=1$.   It is well known that the Hilbert space of the Luttinger model maps exactly to the Hilbert space of non-interacting bosons, and that the fermion bilinear density operator maps to the boson operator \cite{haldane1981luttinger, Giamarchi2004}.   Thus, the vanishing of the higher order connected density correlations is simply a reflection of Wick's theorem for the non-interacting bosons.  Eq. \eqref{sm>d+1} therefore appears to be a higher dimensional generalization of that structure.   However, there is a subtle difference.   Under bosonization, the bosons are non-interacting only when the fermion dispersion is perfectly linear (the Luttinger model).   Nonlinearities in the fermion dispersion - even for non-interacting fermions - lead to interactions among the bosons \cite{haldane1981luttinger}.  In general, for interacting bosons, the higher order connected boson correlations will {\it not} vanish, yet we have shown for {\it non-interacting fermions} the density correlations {\it do} vanish.   The resolution is that for non-interacting fermions, curvature in the fermion dispersion leads to corrections in the density operator.   Those corrections, in combination with the interactions between the bosons lead to an exact cancellation in the higher order connected density correlations computed using bosonization.   This cancellation is specific to non-interacting fermions.
Thus, while bosonization in $D=1$ remains exact when interactions are incorporated in the Luttinger model, the structure of the equal time density correlations outlined in this paper for $D>1$ appear to be a feature that applies only to non-interacting fermions.  A theory of nonlinear bosonization has recently been introduced for dimension $D>1$ Fermi liquids, which incorporates both interactions between the bosons and nonlinear corrections to the density operators \cite{Son2022}.   A similar exact cancellation must arise in that theory for the equal time density correlations when applied to non interacting fermions.

We have proposed two methods for measuring these universal correlations experimentally.   For fermionic atoms in a quantum gas microscope, we argued that for a finite system with of order 100 atoms the universal correlations are apparent in the Fourier transform of the real space density correlations.   For a solid state system, we proposed measuring correlations in the speckle pattern produced by scattering X-rays from an electron gas.   While achieving the required angular and temporal resolution presents a technical challenge, our hope is that the potential for measurement of a fundamental quantity provides motivation.
  
In this work we have not addressed the effect of interactions among the fermions.
For an atomic gas, the interactions among fermions can be controlled, so the prospect of studying a non-interacting (or very weakly interacting) gas of fermionic atoms is within reach.   In a solid state system, interactions among electrons are more difficult to control, but in principle they can be modified by the dielectric environment.   In any event, it remains an important problem to understand the effect that interactions will have on the universal density correlations.   Or to turn it around:  what can measuring the density correlations in a Fermi gas teach us about the interactions?

In one dimension, the answer is known.   For an interacting Fermi gas in one dimension the long wavelength density correlations can easily be computed using bosonization.   For a single channel (spinless) Luttinger liquid \cite{Giamarchi2004}, 
\begin{equation}
    s_2(q) = \frac{|q|}{2\pi}K,
\end{equation}
where $K$ is the dimensionless Luttinger parameter that characterizes the strength of the long wavelength forward scattering interactions among the fermions.    Thus, the $|q|$ singularity, which reflects long ranged real space correlations remains, but its magnitude is modified by the interactions.

We anticipate that a similar modification will arise in higher dimensions.   The analogs of the Luttinger parameter in a higher dimensional Fermi liquid are the Landau parameters, which likewise characterize the strength of the long wavelength forward scattering interactions among the fermions.   It will therefore be of interest to determine the manner in which the Landau parameters modify the universal density correlations in a two or three dimensional Fermi liquid.   We will leave that analysis for future work.

\begin{acknowledgments}
We thank Paul Heiney and Nick Read for helpful discussions.
This work was supported by a Simons Investigator Grant to C.L.K.\ from the Simons Foundation. 
\end{acknowledgments}

\appendix

\section{Proof of Eq. \eqref{sd+1} for general $D$}\label{appendix: general D}

In this appendix we present a proof of Eq. \eqref{sd+1} that unifies our arguments for dimensions $D=1,2$ and $3$ and generalizes the result to arbitrary $D$.   For $M=D+1$ the correlation function may be written as
\begin{equation}
s_{D+1}(\{{\bf q}_a\}) = \int \frac{d^D{\bf q}_{D+1}}{(2\pi)^D}
\langle  \prod_{a=1}^{D+1} \left( \int \frac{d^D{\bf k}_a}{(2\pi)^D} c_{{\bf k}_a}^\dagger c_{{\bf k}_a + {\bf q}_a} \right) \rangle_c.
\end{equation}
We will begin by evaluating the connected correlation function using Wick's theorem.  We will then relate this result to a triangulation of the $D$-dimensional Fermi sea based on a lattice in momentum space generated by $\{{\bf q}_a\}$.

\subsection{Evaluation via Wick's theorem}

The first step is to evaluate the expectation value
\begin{equation}
\langle c^\dagger_{{\bf k}_1} c_{{\bf k}_1+{\bf q}_1}c^\dagger_{{\bf k}_2} c_{{\bf k}_2+{\bf q}_2}
...c^\dagger_{{\bf k}_{D+1}} c_{{\bf k}_{D+1}+{\bf q}_{D+1}}\rangle_c
\label{wickconnected}
\end{equation}
using Wick's theorem.    There will be $D!$ terms in the connected correlation function, in which $c$'s and $c^\dagger$'s are contracted in different pairings.  These can be  specified by a permutation of $\{1,...,D\}$, which we represent as a string $[a] \equiv a_1a_2...a_D$, that specifies that $c_{{\bf k}_{a_i}+{\bf q}_{a_i}}$ is contracted with $c_{{\bf k}_{a_{i+1}}}^\dagger$.   In addition, $c_{{\bf k}_{D+1} +{\bf q}_{D+1}}$ is contracted with $c^\dagger_{{\bf k}_{a_1}}$.  It will be convenient below to represent the substring consisting of the first $i$ elements of $[a]$ 
as $[a]_i \equiv a_1a_2...a_i$, and to define ${\bf q}_{[a]_i} = {\bf q}_{a_1} + {\bf q}_{a_2} + ...+ {\bf q}_{a_i}$.

We start on the right hand side and contract $c_{{\bf k}_{D+1} + {\bf q}_{D+1}}$ (which we define as $c_{\bf k}$) with $c_{{\bf k}_{a_1}}^\dagger$.   This gives
\begin{equation}
\langle c_{{\bf k}_{a_1}}^\dagger c_{\bf k}\rangle = (2\pi)^D f_{{\bf k}_{a_1}}  \delta({\bf k}_{a_1}-{\bf k}).
\label{f0}
\end{equation} 
Next contract $c_{{\bf k}_{a_1}+{\bf q}_{a_1}}$ with $c_{{\bf k}_{a_2}}^\dagger$, and so on.   At each step, we obtain 
\begin{equation}
\langle c_{{\bf k}_{a_{i+1}}}^\dagger c_{{\bf k}_{a_i} + {\bf q}_{a_i}}\rangle = (2\pi)^D f_{{\bf k}_{a_{i+1}}}\delta({\bf k}_{a_{i+1}}-{\bf k}_{a_i} - {\bf q}_{a_i})
\label{f1}
\end{equation} 
for $a_{i+1}<a_i$ or
\begin{equation}
\langle c_{{\bf k}_{a_i} + {\bf q}_{a_i}} c_{{\bf k}_{a_{i+1}}}^\dagger \rangle = (2\pi)^D \bar f_{{\bf k}_{a_{i+1}}}\delta({\bf k}_{a_{i+1}}-{\bf k}_{a_i} - {\bf q}_{a_i})
\label{f2}
\end{equation}
for $a_{i+1}>a_i$ (where $\bar f = 1-f$).
After the $D+1$ $\delta$-functions are integrated, there remains a single integral over ${\bf k}$, and we set ${\bf k}_{a_1}={\bf k}$ and ${\bf k}_{a_{i+1}} = {\bf k} + {\bf q}_{[a]_i}$ for $i=1, ..., D$.

The contribution from each permutation $[a]$ will be a product of (\ref{f0}) and terms (\ref{f1}) or (\ref{f2}) for $i=1,...,D$ up to an overall sign that accounts for the antisymmetry of the fermion operators.   This sign can be deduced by considering the sign of the term that contains all $D+1$ $f$'s when the product is multiplied out using $\bar f = 1-f$.   
This is determined by rearranging the $c^\dagger$'s in (\ref{wickconnected}) such that $c^\dagger_{{\bf k}_{a_{i+1}}}$ sits to the left of $c_{{\bf k}_{a_i}+{\bf q}_{a_i}}$, and counting the number of interchanges of fermion operators.   This can be accomplished in three steps: \\

(i) Move all of the $c^\dagger$'s to the left of all of the $c$'s (keeping their order preserved).  

(ii) Rearrange the $c^\dagger$'s according to the permutation $[a]$ by moving $c^\dagger_{{\bf k}_{a_{i+1}}}$ to the position previously occupied by $c^\dagger_{{\bf k}_{a_{i}}}$ for $0\le i\le D$ (with $a_0=a_{D+1}=D+1$). 

(iii) Move the $c^\dagger$'s back to their original positions (keeping their updated order preserved).\\

\noindent Steps (i) and (iii) will involve the same number of interchanges, so they will cancel.   The overall sign is therefore determined by the sign of the permutation $[a]$.  Since $[a]$ represents a cycle of length $D+1$, its parity is $(-1)^{D}$.  

It follows that
\begin{equation}
s_{D+1} = \int \frac{d^D{\bf k}}{(2\pi)^D} \sum_{[a]}f_0 \prod_{i=1}^{D} \left(\theta(a_{i+1}-a_i) - f_{[a]_i}\right),
\label{sD+1int}
\end{equation}  
where we adopt the shorthand $f_0 \equiv f_{\bf k}$ and $f_{[a]_i} \equiv f_{{\bf k} + {\bf q}_{[a]_i}}$.

\subsection{Triangulated Fermi Sea}

We now generalize the notion of a triangulated Fermi sea to $D$ dimensions.   We begin by considering the set of points  ${\mathcal L}({\bf k}_0)$ in momentum space containing a point ${\bf k}_0$, along with all points related by the lattice generated by ${\bf q}_1, ..., {\bf q}_D$:
\begin{equation}
{\bf k}_{\bf n} = {\bf k}_0 + \sum_{a=1}^D n_a {\bf q}_a
\label{kntriangulate}
\end{equation}
for integers $n_a$.   

In general, this lattice will not fit inside the Brillouin zone unless a multiple of ${\bf q}_a$ is a primitive reciprocal lattice vector.   However, if the ${\bf q}_a$ are {\it rational fractions} of a reciprocal lattice vector, then the lattice will fit into $N$ copies of the Brillouin zone for some (possibly large) integer $N$.   We will be content with establishing our result for ${\bf q}_a$ that are (arbitrarily high order) rational fractions.   In this case, the sums over $\{n_a\}$ should be understood as summing over the distinct points in the $N$-fold expanded Brillouin zone, and we must keep in mind that the Fermi sea is repeated $N$ times.  

For a point ${\bf k}$ in ${\mathcal L}({\bf k}_0)$, a string $[a] = a_1a_2..a_d$, and a collection of integers  $ 0< i_1 < i_2 < ... < i_d < D+1$, we define the {\it  $d$-simplex},
\begin{equation}
\sigma^{\bf k}_{[a]_{i_1}[a]_{i_2}...[a]_{i_d}}  = ({\bf k}, {\bf k}+{\bf q}_{[a]_{i_1}}, {\bf k}+{\bf q}_{[a]_{i_2}}, ... , {\bf k}+{\bf q}_{[a]_{i_d}}),
\end{equation}
which is defined by its set of $d+1$ corners.
Associated with each ${\bf k} \in {\mathcal L}$ there are $D!$ $D$-simplexes,
$\sigma^{\bf k}_{[a]_1[a]_2...[a]_D}$, corresponding to the $D!$ permutations $[a]$, that together make up a unit cell of ${\mathcal L}$ given by the $D$-dimensional parallelepiped at ${\bf k}$ generated by  ${\bf q}_{1,...,D}$. 
The set of all $d$-simplexes for $d=0,1,2, ..., D$  for all ${\bf k}$ in ${\mathcal L}$ form a {\it simplicial complex} ${\mathcal F}({\bf k}_0)$ that defines a {\it triangulation} of the ($N$-fold) Brillouin zone. These are the building blocks for our triangulation of the Fermi sea.   

The set of all simplexes in $\mathcal{F}({\bf k}_0)$, whose corners are all contained in inside the Fermi sea defines a simplicial complex ${\mathcal F}_{NF}({\bf k}_0)$.   This defines an approximate triangulation of the ($N$-fold) Fermi sea.   The Euler characteristic of this simplicial complex can be evaluated by the Euler-Poincar\'e theorem \cite{Nash1988, Nakahara1990},
\begin{equation}
\chi_{NF}({\bf k}_0) = \sum_{d = 0}^D (-1)^d \mathcal{N}_d \big({\mathcal F}_{NF}({\bf k}_0)\big),
\label{EulerPoincare}
\end{equation}
where $\mathcal{N}_d$ is the number of $d$-simplexes in ${\mathcal F}_{NF}({\bf k}_0)$.

We expect that for sufficiently small ${\bf q}_a$ this approximate triangulation should faithfully represent the topology of the Fermi sea, though for large ${\bf q}_a$ it will not.   In the following section we will assume that $\chi_{NF}({\bf k}_0)$ is independent of ${\bf k}_0$ and is related to the Euler characteristic of the Fermi sea.
We will show that Eq. \eqref{sD+1int} precisely evaluates Eq. \eqref{EulerPoincare}, which allows us to express $s_{D+1}$ in terms of $\chi_F$.   In Appendix \ref{appendix:bound} we will consider the range of validity of the approximate triangulation, and present a bound on how small ${\bf q}_a$ should be.

\subsection{Evaluation of $s_{D+1}$}

We now evaluate Eq. \eqref{sD+1int}.   We begin by  expressing the ${\bf k}$-integral as $1/N$ times the integral over the $N$-fold Brillouin zone, which using Eq. \eqref{kntriangulate} can  be expressed as an integral of ${\bf k}_0$ over the parallelepiped $P$ formed by ${\bf q}_{1,...,D}$ times a sum over the lattice ${\mathcal L}({\bf k}_0)$. Namely,
\begin{equation}
s_{D+1}(\{{\bf q}_a\})  = \frac{1}{N}\int_P \frac{d^D{\bf k}_0}{(2\pi)^D} C({\bf k}_0),
\label{sd+1c}
\end{equation}
with
\begin{equation}
C({\bf k}_0) = \sum_{{\bf k} \in {\mathcal L}({\bf k}_0)} \sum_{[a]} f_0 \prod_{i=1}^D\left( \theta(a_{i+1}-a_i)-f_{[a]_i}\right).
\label{ck0}
\end{equation}
We will show that $C({\bf k}_0)$ precisely evaluates $\chi_{NF}({\bf k}_0)$ by expanding the product in Eq. \eqref{ck0}.  

First consider the term in which all $D$ terms in the product involve $f_{[a]_i}$, which corresponds to $D$-simplexes.  It is clear from Eq. \eqref{ck0} that each of the $D!$ permutations $[a]=a_1a_2...a_D$ involves
\begin{equation}
 f_0 \prod_{i=1}^D f_{[a]_i} = 
\left\{\begin{array}{ll}
1, & \sigma^{\bf k}_{[a]_1[a]_2...[a]_D} \in {\mathcal F}_{NF}({\bf k}_0)\\
0, & {\rm otherwise}.
\end{array}
\right. 
\end{equation}
When summed over ${\bf k}\in {\mathcal L}({\bf k}_0)$, these terms give $(-1)^D \mathcal{N}_D\big({\mathcal F}_{NF}({\bf k}_0)\big)$.

Next, for $d<D$ consider the terms with $d$ $f_{[a]_i}$'s, corresponding to $d$-simplexes that are $d$-faces of the $D$-simplex associated with $[a]$.  By choosing integers $\{i_1, i_2,...,i_d\}$ in ascending order out of $\{1,2,...,D\}$, such a simplex is constructed as $\sigma^{\bf k}_{[a]_{i_1}[a]_{i_2}...[a]_{i_d}}$.  The corresponding term in the expansion of Eq. \eqref{ck0} is
\begin{equation}
(-1)^d f_0 f_{[a]_{i_1}} f_{[a]_{i_2}}...f_{[a]_{i_d}}.
\end{equation}
Since a $d$-face is shared by multiple $D$-simplexes, $\sigma^{\bf k}_{[a]_{i_1}[a]_{i_2}...[a]_{i_d}}$ can be built from various permutations $[a]$ and different choices of indices $\{i_1,i_2,...,i_d\}$. 
Nonetheless, each $d$-simplex is counted precisely once in Eq. \eqref{ck0} because  the $\theta$-functions there imply that
\begin{equation}
a_{i_n} > a_{i_n-1} > a_{i_n - 2} > ... > a_{i_{n-1}+1}
\end{equation}
for $n = 1,2,...,d+1$ (Here we denote $i_0 = 0$, $i_{D+1}=D+1$).  These constraints uniquely fix the choice of permutation $[a]$ and the indices $\{i_1,i_2,...,i_d\}$ for each $d$-simplex.  For example, for $D=2$, $d=1$, the term $f_0 f_{12}$ in Eq. \eqref{s3sum} corresponds to a 1-simplex that is a face of either of the 2-simplexes represented by $[a] = 12$ or $[a]=21$, with $i_1=2$.  In Eq. \eqref{ck0} this term comes with $\theta(a_2-a_1)$, which selects $[a]=12$.  Similarly, for $D=3$, $d=1$ the term $f_0 f_{123}$ in Eq. \eqref{s4sum} corresponds to a $1$-simplex that is an edge of $6$ $3$-simplexes $a_1a_2a_3$, but $\theta(a_2-a_1)\theta(a_3-a_2)$ selects $123$.

When summed over ${\bf k}\in {\mathcal L}({\bf k}_0)$, each $d$-simplex in ${\mathcal F}_{NF}({\bf k}_0)$ is counted once, so these terms contribute $(-1)^d \mathcal{N}_d\big({\mathcal F}_{NF}({\bf k}_0)\big)$.
Adding up the contributions for all of the $d$-simplexes, we conclude that $C({\bf k}_0)$ evaluates the Euler-Poincar\'e formula for the Euler characteristic,
\begin{equation}
C({\bf k}_0)  
= \sum_{d=0}^D (-1)^d \mathcal{N}_d({\mathcal F}_{NF}\big({\bf k}_0)) \equiv \chi_{NF}({\bf k}_0\big).
\label{ck00}
\end{equation}
Provided we assume that ${\bf q}_a$ are small enough such that the triangulation ${\cal F}_{NF}({\bf k}_0)$ faithfully represents the topology of the Fermi sea, $\chi_{NF}({\bf k}_0)$ will be independent of ${\bf k}_0$.

Eq. \eqref{ck00} can alternatively be written as
\begin{equation}
    \chi_{NF}({\bf k}_0) = \sum_{\sigma \in \mathcal{F}({\bf k}_0)} (-1)^{d_\sigma}  \prod_{\{\bk_i\in\sigma\}} f_{\bk_i},
    \label{chinf f}
\end{equation}
where $d_\sigma$ is the dimensionality of simplex $\sigma$.  It can be observed that the sums in Eqs. \eqref{s3sum} and \eqref{s4sum} have this form.

All that is left is to account for the fact that for $N>1$ $\chi_{NF}$ will be equal to the Euler characteristic of the $N$-fold Fermi sea, rather than the actual Fermi sea.   A simple way to relate the two is to use Morse theory to express the Euler characteristic as a sum over the critical points in the energy $E({\bf k})$, which can be regarded as a Morse function.   Clearly, the $N$-fold Fermi sea will have $N$ copies of every critical point in the original Fermi sea.   It follows that $\chi_{NF} = N \chi_F$. 

We thus conclude that $C({\bf k}_0) = N \chi_F$.   The integral over ${\bf k}_0$ in (\ref{sd+1c}) simply gives the volume of $P$, which is given by $\abs{{\rm det}[{\mathbb Q}]}$, where ${\mathbb Q}$ is the $D\times D$ matrix formed by ${\bf q}_{1,...,D}$.  Eq. \eqref{sd+1} follows.

\section{Bound on ${\bf q}_a$}
\label{appendix:bound}

In this appendix we consider the range of ${\mathbb Q} = [{\bf q}_1, ...,{\bf q}_D]$ for which Eq. \eqref{sd+1} is exact.   This requires us to assess the validity of our assertion that $\chi_F$ of the Fermi sea is faithfully represented by $\chi_{NF}({\bf k}_0)$ of the approximate triangulation of the Fermi sea based on the lattice ${\mathcal L}({\bf k}_0)$, which is generated by ${\mathbb Q}$.   If the ${\bf q}_a$'s are small, such that the mesh of points is fine on the scale of the size $k_F$ of the Fermi sea, then it is natural to expect that the triangulation gets the topology right.   However, when ${\bf q}_a$ is larger than $2k_F$, it is possible that the lattice misses the Fermi sea entirely.   Moreover, even when the ${\bf q}_a$'s are small, our approach breaks down if the ${\bf q}_a$'s are linearly dependent (so that ${\rm det}[{\mathbb Q}] = 0)$, since in that case ${\mathcal L}$ will not fill the Brillouin zone.  In fact, when $|{\rm det}[{\mathbb Q}]|$ is small, the criterion for how small ${\bf q}_a$ must be becomes more restrictive, and depends on the curvature of the Fermi surface.    

Our strategy for identifying the regime in which Eq. \eqref{sd+1} is exact is to demand that $\chi_{NF}({\bf k}_0)$ is independent of ${\bf k}_0$.   Thus, it can not change if a point in ${\mathcal L}({\bf k}_0)$ passes from inside to outside the Fermi surface as ${\bf k}_0$ is varied.   It is thus sufficient to set ${\bf k}_0 = {\bf k}$, for an arbitrary point ${\bf k}$ on the Fermi surface, and ask whether $\chi_{NF}({\bf k}_0)$ depends on whether the point ${\bf k}_0$ is just inside or just outside the Fermi sea.  Consider
\begin{equation}
\Delta \chi({\bf k}) = \chi_{NF}({\bf k} - \epsilon \hat n_{\bf k}) - \chi_{NF}({\bf k} + \epsilon\hat n_{\bf k})
\end{equation}
for $\epsilon\rightarrow 0$, and $\hat n_{\bf k} \propto \nabla_{\bf k} E({\bf k})$ is a unit normal to the Fermi surface.
Since $\chi_{NF}({\bf k}_0)$ counts the numbers of $d$-simplexes in the Fermi sea, it is clear that the terms in Eq. \eqref{chinf f} that depend on $f_{\bf k}$ will only involve the {\it neighbors} of ${\bf k}$, ${\mathcal L}^{\rm neighbors}({\bf k})$, defined as the points in ${\mathcal L}({\bf k})$ that are connected to ${\bf k}$ by a $1$-simplex in ${\mathcal F}({\bf k})$.   We will denote the neighbors of ${\bf k}$ as ${\bf k}_h \equiv {\bf k}+{\bf q}_h$ and ${\bf k}_{\overline h} \equiv {\bf k}-{\bf q}_h$, where $h = \{h_1,...,h_d\}$ is a subset of $\{1,...,D\}$, and ${\bf q}_h = \sum_{i=1}^d {\bf q}_{h_i}$, for $d=1,2,...,D$.

It is straightforward to show that
\begin{equation}
\Delta \chi({\bf k}) = \chi^{\rm neighbors}_F({\bf k}) -1 ,
\end{equation}
where
\begin{equation}
\chi^{\rm neighbors}_F({\bf k}) = \sum_{d=0}^D (-1)^d \mathcal{N}_d[{\mathcal F}^{\rm neighbors}_F({\bf k})],
\end{equation}
and ${\mathcal F}_F^{\rm neighbors}({\bf k})$ is a simplicial complex defined as the subset of ${\mathcal F}({\bf k})$ for which every vertex ${\bf k}_h$ of every simplex is in ${\mathcal L}^{\rm neighbors}({\bf k})$ and is inside the Fermi sea.  Thus, the criterion for $\chi_{NF}({\bf k}_0)$ to be independent of ${\bf k}_0$ is that 
\begin{equation}
\chi_F^{\rm neighbors}({\bf k}\in S_F)=1
\label{criterion}
\end{equation}
for every point ${\bf k}$ on the Fermi surface, $S_F$.

\begin{figure}
    \centering
    \includegraphics[width=3.2in]{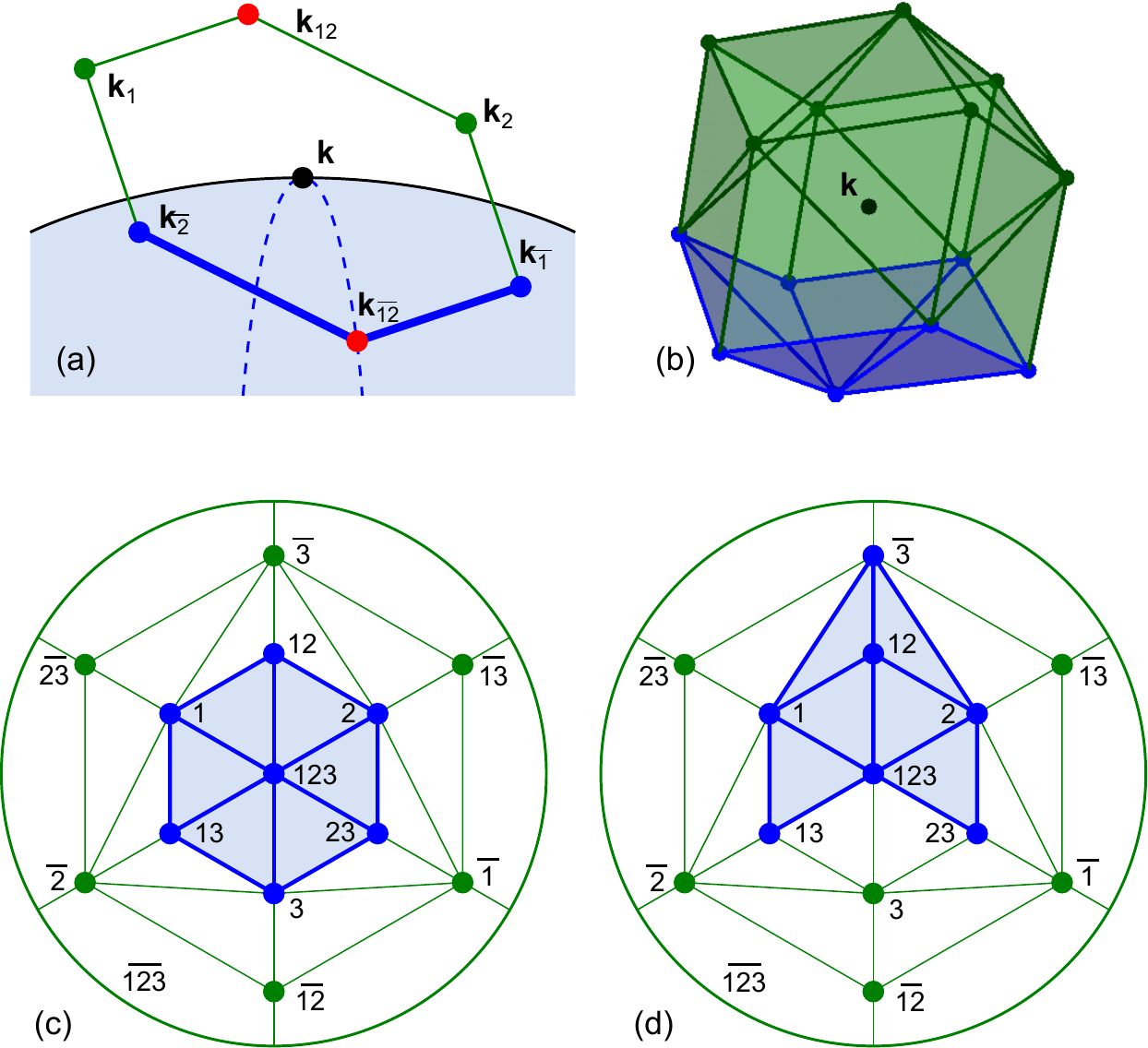}
    \caption{Representative illustration of ${\mathcal L}^{\rm neighbors}({\bf k})$, i.e. the set of points labeled as $\bk_{h,\overline{h}}$ (or $h, \overline{h}$ in short) that surrounds $\bk$, and ${\mathcal F}_F^{\rm neighbors}({\bf k})$, i.e. the corresponding simplicial complex inside the Fermi sea labeled in blue. Here $\bk$ is a generic point on the Fermi surface. (a) for $D=2$, where the red points impose bounds on the Fermi surface curvature to ensure the exactness of Eq. \eqref{s3chiF}; (b) for $D=3$, where (c,d) are projections showing the possible configurations of ${\mathcal F}_F^{\rm neighbors}$.}
    \label{fig:neighbors}
\end{figure}

For $D=2$, the 6 neighbors of ${\bf k}$ form a (distorted) hexagon, shown in Fig. \ref{fig:neighbors}(a).   
For $D=3$, the 14 neighbors of ${\bf k}$ form the vertices of a (distorted) rhombic dodecahedron (Fig. \ref{fig:neighbors}(b)).  
To aid visualization for $D=3$, Figs. \ref{fig:neighbors}(c,d) show a projection of the sphere containing the neighbors onto a disk, with one of the neighbors ${\bf k}_h$ at the center of the disk, and its partner ${\bf k}_{\overline{h}}$ represented by the boundary of the disk.   
Generically, when the Fermi surface is flat (or nearly flat on the scale of  ${\bf q}_h$), ${\bf k}_h$ and ${\bf k}_{\bar h}$ will be on opposite sides of the Fermi surface, so the Fermi surface divides the neighbors in half.    For $D=2$ ($D=3$) the 3 (7) neighbors inside the Fermi sea are shown in blue.   Depending on the orientation $\hat n_{\bf k}$ of the Fermi surface relative to the ${\bf q}_h$'s, different combinations of neighbors will be included, but the pattern always resembles Fig. \ref{fig:neighbors}(a) for $D=2$ and either Fig. \ref{fig:neighbors}(c) or Fig. \ref{fig:neighbors}(d) for $D=3$.  Clearly, in these cases $\chi_F^{\rm neighbors}({\bf k}) = 1$.

For general $D$, the $2(2^D-1)$ neighbors of ${\bf k}$ form a triangulation of a sphere $S^{D-1}$.   
The hemisphere inside the Fermi sea will  have the topology of a $D-1$ dimensional ball, with $\chi^{\rm neighbors}_F({\bf k}) = 1$, so that Eq. \eqref{criterion} is satisfied.   However, this criterion will break down if the Fermi surface deviates too strongly from being flat, or equivalently, if ${\bf q}_h$'s are too large.

In general, the range of ${\mathbb Q}$ for which Eq. \eqref{sd+1} will be exact will depend on the detailed shape of the Fermi surface, which can in principle be arbitrarily complicated.   Rather than attempting a completely general result, we will assume that the Fermi surface is smooth, and is locally characterized by a curvature, so that near every point ${\bf k}$ on the Fermi surface we can express the nearby Fermi surface points as ${\bf k} + {\bf q}$, where
\begin{equation}
\hat n_{\bf k}\cdot{\bf q}  = {\bf q} \cdot {\mathbb C}_{\bf k} \cdot {\bf q}.
\label{kf(q)}
\end{equation}
Here $\hat n_{\bf k}$ is normal to the Fermi surface at ${\bf k}$, and  ${\mathbb C}_{\bf k}$ is a rank $D-1$ matrix (satisfying $\hat n_{\bf k}\cdot {\mathbb C}_{\bf k} = {\mathbb C}_{\bf k}\cdot\hat n_{\bf k}=0$) characterizing the curvature of the Fermi surface at ${\bf k}$.   For a spherical Fermi surface of radius $k_F$, ${\mathbb C} = (2k_F)^{-1}({\mathbb I}-\hat n_{\bf k}\hat n_{\bf k})$.   Moreover, we will assume that ${\bf q}_h$ is already small enough, such that Eq. \eqref{kf(q)} is accurate for every ${\bf k}$ on the Fermi surface at the scale of ${\bf q}_h$.   Small higher order corrections to Eq. \eqref{kf(q)} will lead to small higher order corrections to our bound, but they will not diminish the exactness of Eq. \eqref{sd+1} when the bound is satisfied.   We will develop our bound for the special cases $D=2$ and $D=3$.   This analysis suggests an algorithm for determining a non-trivial bound, which we  conjecture is valid for all $D$.

\subsection{Bound for $D=2$}

In two dimensions, it can be seen from Fig. \ref{fig:neighbors}(a) that for a flat Fermi surface, three of the six neighbors of a point ${\bf k}$ on the Fermi surface will be inside the Fermi sea, forming a simplicial complex ${\mathcal F}_F^{\rm neighbors}(\bk)$ with $\chi_F^{\rm neighbors}({\bf k})=1$.  Assuming Eq. \eqref{kf(q)}, ${\mathcal F}_F^{\rm neighbors}$ will be unchanged for a curved Fermi surface, provided $|\hat n_{\bf k}\cdot{\bf q}_h| > |{\bf q}_h \cdot {\mathbb C} \cdot {\bf q}_h|$ for all neighbors ${\bf k}_h \in {\mathcal F}^{\rm neighbors}_F$.   However, this criterion is too restrictive because near a point on the Fermi surface where $\hat n_{\bf k} \cdot {\bf q}_h =  0$, the criterion is always violated.   However, it can be seen from Fig. \ref{fig:neighbors}(a) that removing ${\bf k}_{\overline{1}}$ or ${\bf k}_{\overline{2}}$ from 
${\mathcal F}_F^{\rm neighbors}$, or adding ${\bf k}_1$ or ${\bf k}_2$ to ${\mathcal F}_F^{\rm neighbors}$, will not change $\chi_F^{\rm neighbors}({\bf k})$.    On the other hand, if the curvature is so high that ${\bf k}_{\overline{12}}$ (or ${\bf k}_{12}$) crosses the Fermi surface (as shown by the dashed line in Fig. \ref{fig:neighbors}(a)), then $\chi_F^{\rm neighbors}({\bf k})=0$, so that $\Delta \chi({\bf k}) = 1$.   

We can obtain a non-trivial bound by removing from the list of neighbors ${\mathcal L}^{\rm neighbors}({\bf k})$ the points that do not matter.   Note that these are precisely the points that could first intersect a flat Fermi surface passing through ${\bf k}$ (perpendicular to $\hat n$) as $\hat n$ is rotated away from its initial value $\hat n_{\bf k}$.   Clearly, for $D=2$, the only points left are the ones that are furthest away from the flat Fermi surface:
\begin{equation}
|\hat n_{\bf k}\cdot {\bf q}_{h^*}| = \max_h |\hat n_{\bf k}\cdot {\bf q}_h|.
\end{equation}
Thus, in Fig. \ref{fig:neighbors}(a), ${\bf q}_{h^*} = \pm {\bf q}_{12} = \mp {\bf q}_3$.   The non-trivial bound is thus,
\begin{equation}
\left|\frac{{\bf q}_{h^*}\cdot {\mathbb C}_{\bf k} \cdot {\bf q}_{h^*}}{\hat n_{\bf k}\cdot {\bf q}_{h^*}}\right| < 1
\label{bound2}
\end{equation}
for all ${\bf k}\in S_F$.   Provided ${\bf q}_1$ and ${\bf q}_2$ are linearly independent (${\rm det}[{\mathbb Q}] \ne 0$), this bound can be satisfied for sufficiently large $k_F$ (or small $|{\bf q}_a|$).

Note that there are two ways in which this bound can be violated.   If we fix the angles between ${\bf q}_a$, then (\ref{bound2}) will be violated when the magnitudes $|{\bf q}_a|$ are turned up too large.   On the other hand, if we fix the magnitudes $|{\bf q}_a|$, then the bound will be violated if ${\bf q}_a$'s become nearly parallel to one another (so that $|{\rm det}{\mathbb Q}|= |{\bf q}_1 \times {\bf q}_2| \ll |{\bf q_1}||{\bf q}_2|$).   In this case, there can be a point on the Fermi surface where $\hat n_{\bf k} \cdot \bq_{h^*} \approx 0$, so (\ref{bound2}) is violated.

For a circular Fermi surface of radius $k_F$, a bound that is exact even when $\abs{{\bf q}_a}/k_F = {\cal O}(1)$ can be found by replacing (\ref{bound2}) by $({q}_{h^*}^\perp)^2 + (k_F - q_{h^*}^\parallel)^2 < k_F^2$, where $\parallel$ ($\perp$) indicate components parallel (perpendicular) to $\hat n_\bk$.   For ${\bf q}_{h^*} = \pm {\bf q}_a$, the bound is determined by the points on the Fermi surface with $\hat n_{\bf k} \perp {\bf q}_b$ for $b\ne a \in \{1,2,3\}$.  
This, in turn, implies the bound can be written as
\begin{equation}
|{\bf q}_a| < 2 k_F \sin \theta_{ab}
\label{circlebound}
\end{equation}
for all $a \ne b \in \{ 1,2,3\}$.  Here $\theta_{ab}$ is the angle between ${\bf q}_a$ and ${\bf q}_b$.   It can be checked that if $\sin\theta_{ab} \ll 1$ (so that the bound applies for $|{\bf q}_a| \ll k_F$), then (\ref{bound2}) and (\ref{circlebound}) agree.

Thus, when $\theta_{ab}$ are of order unity, Eq. \eqref{sd+1} will be exact when the magnitudes $|{\bf q}_a|$ are smaller than a cutoff of order $k_F$.   But when the ${\bf q}_a$'s are nearly parallel, the condition on $|{\bf q}_a|$ is more restrictive.   

\begin{figure}
    \centering
    \includegraphics[width=3.2in]{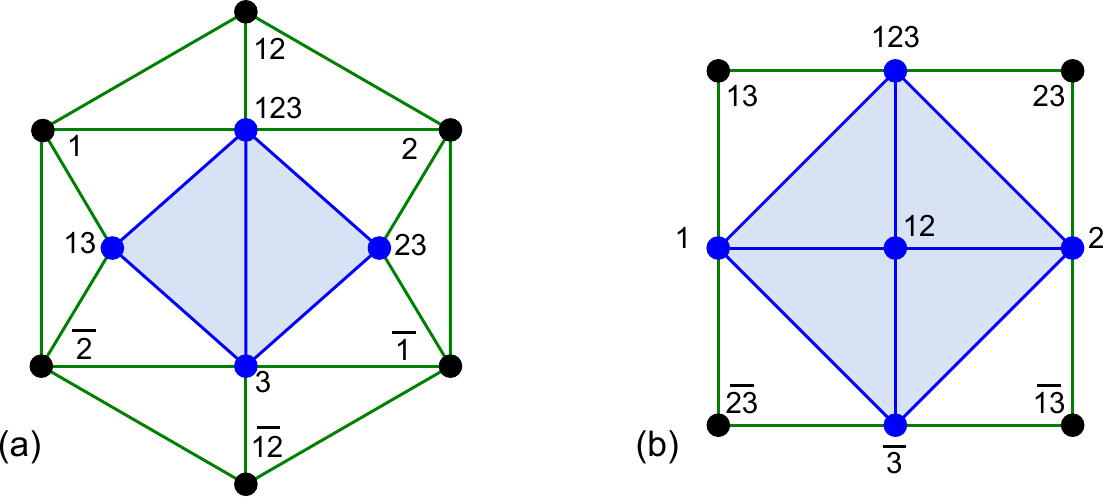}
    \caption{Neighbors in $D=3$. The black points are the marginal neighbors of $\bk$, which do not affect $\chi_F^{\rm neighbors}$ irrespective of whether they are included in the Fermi sea. The blue points belong to ${\mathcal L}^{* \rm neighbors}({\bf k})$, which are relevant for imposing bounds on the Fermi surface curvature to ensure the exactness of Eq. \eqref{s4chiF}. (a) and (b) correspond to Fig. \ref{fig:neighbors}(c) and (d), respectively.}
    \label{fig:3dneighbors}
\end{figure}

\subsection{Bound for $D=3$}

The procedure for $D=3$ is similar to that for $D=2$, above, except the counting is a bit more complicated.   
A bound similar to (\ref{bound2}) can be established, provided we remove from ${\mathcal L}^{\rm neighbors}({\bf k})$ the marginal neighbors for which $\bq_{h} \cdot \hat n_\bk$
could be arbitrarily small, which we will argue will not affect $\chi_F^{\rm neighbors}$.   
These are the neighbors that could first intersect a flat Fermi surface perpendicular to $\hat n$ passing through ${\bf k}$ as $\hat n$ is rotated away from $\hat n_{\bf k}$.   The marginal neighbors can be identified by ordering the neighbors according to
\begin{equation}
|\hat n_{\bf k}\cdot{\bf q}_{h_1} | < |\hat n_{\bf k}\cdot {\bf q}_{h_2}| < ... ,
\end{equation}
and selecting the smallest two from this list, along with any linear combinations of those two that are also neighbors.   Depending on the orientation of $\hat n_{\bf k}$ there could be 6 or 4 marginal neighbors, which all lie in the same plane containing ${\bf k}$.   These are shown in black in Fig. \ref{fig:3dneighbors}(a,b), and by inspecting these pictures, it is clear that the Euler characteristic $\chi_F^{\rm neighbors}$ is independent of whether any of the marginal neighbors are included in the Fermi sea or not.   This can be checked by explicitly writing $\chi_F^{\rm neighbors}({\bf k})$ in terms of $f_h \in \{0,1\}$ in a form analogous to Eq. \eqref{s3sum} and observing that the result is independent of $f_h$ if ${\bf k}_h$ is a marginal neighbor.
Thus, the marginal neighbors do not need to be included in the bound.

We thus define ${\mathcal L}^{* \rm neighbors}({\bf k})$ to be the points in ${\mathcal L}^{\rm neighbors}$ with the marginal neighbors removed.   The bound then becomes
\begin{equation}
\left|\frac{{\bf q}_{h^*} \cdot {\mathbb C}_{\bf k} \cdot {\bf q}_{h^*}}{\hat n_{\bf k}\cdot{\bf q}_{h^*}}\right| < 1
\label{bound3}
\end{equation}
for all ${\bf k} \in S_F$ and for all ${\bf k}_{h^*} \in {\mathcal L}^{*\rm neighbors}$.   Provided $|{\rm det}[{\mathbb Q}]|>0$, we are guaranteed that $|\hat n_{\bf k}\cdot {\bf q}_{h^*}| >0$ for all ${\bf k}$ and $h^*$, so that the bound is non-trivial, and Eq. \eqref{sd+1} is exact over a finite range of ${\bf q}_a$.

\subsection{Conjectured bound for all $D$}

For general $D$, characterizing ${\mathcal L}^{\rm neighbors}({\bf k})$ and ${\mathcal F}_F^{\rm neighbors}({\bf k})$ becomes more complicated, and we have not proven the general case.   Nonetheless, we conjecture that the algorithm introduced above for $D=3$ can be applied for $D>3$:

\begin{itemize}

\item[1.]  For each point ${\bf k}$ on the Fermi surface determine the set of neighbors ${\cal L}^{\rm neighbors}({\bf k})$.

\item[2.]  Identify the marginal neighbors by choosing the $D-1$ linearly independent neighbors in ${\mathcal L}^{\rm neighbors}({\bf k})$ with the smallest values of $|\hat n_{\bf k}\cdot {\bf q}_h|$, along with all linear combinations in ${\mathcal L}^{\rm neighbors}({\bf k})$.  Define ${\mathcal L}^{*\rm neighbors}$ as the points in ${\mathcal L}^{\rm neighbors}({\bf k})$ with the marginal points removed.

\item[3.]  Require (\ref{bound3}) for all ${\bf k}\in S_F$ and ${\bf k}_{h^*} \in {\mathcal L}^{*\rm neighbors}$.

\end{itemize}

\end{document}